\documentclass[twocolumn,showpacs,amsmath,amssymb,prb,superscriptaddress,reprint]{revtex4}

\usepackage{graphicx}
\usepackage{bm}

\begin{document}

\title{Vortex matter in low dimensional systems with proximity induced superconductivity}

\author{ N.B.\ Kopnin } \affiliation{ O.V. Lounasmaa Laboratory,
Aalto University, P.O. Box 15100, 00076
Aalto, Finland} \affiliation{ L.~D.~Landau Institute for
Theoretical Physics, 117940 Moscow, Russia}
\author{ I.M.\ Khaymovich}
\affiliation{Institute for Physics of Microstructures, Russian
Academy of Sciences, 603950 Nizhny Novgorod, GSP-105, Russia }
\author{A.S.\ Mel'nikov}
\affiliation{Institute for Physics of Microstructures, Russian
Academy of Sciences, 603950 Nizhny Novgorod, GSP-105, Russia }

\date{\today}

\begin{abstract}
We study theoretically the vortex matter structure in
low dimensional (LD) systems with superconducting order induced by
proximity to a bulk superconductor. We analyze the effects of microscopic coupling mechanisms between the two systems and
the effects of possible mismatch in the band structures of these materials on the energy spectrum of vortex-core electrons. The unusual structure of vortex cores is discussed in the context of recent tunneling microscopy/spectroscopy experiments.
\end{abstract}

\maketitle

\section{Introduction.}\label{sec-intro}

The induced superconducting order attracts considerable interest of both theorists and experimentalists for many decades starting
from the seminal works on the proximity effect.\cite{deGennes-book,McMillan}
Recently, one sees a revival of this interest in connection with the growing number of experiments carried out for a variety of
new artificial systems which include two-dimensional electron gas, graphene, semiconducting nanowires and carbon nanotubes, topological insulators, etc. Exotic electronic properties of these systems \cite{been_RMP,CastroNeto08,nanotubes_Bouchiat,nanotubes_review,topoins} can cause quite unusual manifestations of the proximity effect.
Superconducting characteristics of such low-dimensional (LD) systems can differ strongly from those in the bulk. Thus the experiments on proximity induced superconductivity provide a unique possibility to manipulate the basic properties of the superconducting state. Control of superconducting characteristics can be realized by changing the doping level through the gate potential, which creates, e.g., new types of tunable Josephson devices.\cite{Graphene-SGS} Unconventional gap potential induces, in turn, unusual quasiparticle (QP) states both in homogeneous and in nonuniform superconducting phases. For LD systems with a nontrivial topological structure one can possibly realize the QP modes with specific symmetries of the electron and hole wave functions at the Fermi level that describe the so-called Majorana fermions in condensed matter.\cite{TI_S_prox_Majorana,TI_S_review}

\begin{figure}[t]
\includegraphics[width=0.75\linewidth]{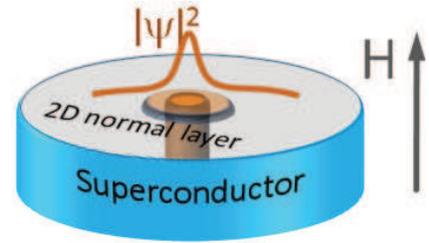}
\caption{(Color online) Sketch of the 2D layer with multiple core vortex structure induced by a bulk type II superconductor in the vortex state. Two scales of the induced vortex are schematically depicted by orange disks in 2D layer and cylinder in the bulk superconductor. }
\label{fig-vort-sketch}
\end{figure}

A standard way of studying the QP states in systems with a complicated superconducting order is to look at the effects of applied magnetic field on the structure of the mixed state. For example, if the bulk electrode is a type-II superconductor (SC) one can study the structure of vortex lines penetrating the electrode and threading also the LD system (see Fig.~\ref{fig-vort-sketch}). It is the goal of this paper to review the basic properties of the vortex matter formed in the LD layer.
Similar problem of the vortex matter in the proximity layers naturally arises
when one faces the challenge of interpreting the scanning tunneling  microscopy/spectroscopy (STM/STS) measurements in superconductors.
Probing the energy and spatial dependencies of the local density of states
(LDOS) by STM/STS \cite{oldSTM} provides information of the spectrum and of the wave functions in the superconducting state.
An important part of this information refers to
the structure of subgap QP states in the magnetic field
bound to the vortex core which are known as the Caroli--de~Gennes--Matricon (CdGM)
states \cite{CdGM}. A fingerprint of these states is the so-called zero-bias anomaly (ZBA) \cite{oldSTM} seen in the STM measurements. Obviously, the intrinsic characteristics of the bound core states can be masked or even hidden by the presence of a thin defect layer at the surface of the bulk SC. In such thin (possibly non-superconducting) surface layer, the superconducting coherence is induced by proximity to the bulk SC. The masking effect of the defect layer is often difficult to distinguish from more exotic explanations based, e.g., on the assumptions of the superconducting gap anisotropy (see \cite{AnisDeltaVortex,Fischer07} and references therein) and multi-component structure of the order parameter \cite{Koshelev,Giubileo01}.
Despite all its simplicity, the model assuming the presence of a defect layer at the sample surface can explain quite a variety of features in the vortex LDOS experimental data and provides an instructive example of the vortex matter in the LD systems with the induced superconducting order.

Instead of considering various phenomenological models of the induced gap potential, in our studies of the vortex matter we rather use
the general microscopic approach developed in Ref.~\onlinecite{KopninMelnikov11} and focus on the physical mechanisms responsible for formation of the particular gap potential and its symmetry.
\begin{figure}[t]
\includegraphics[width=0.5\linewidth]{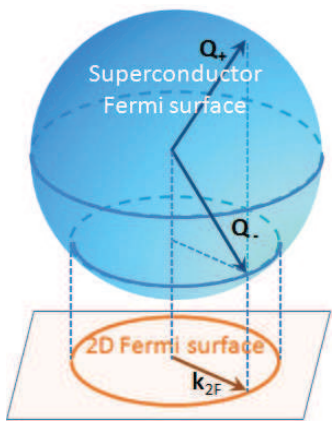}
\caption{(Color online) Matching of Fermi surfaces in 2D layer and in the bulk superconductor in the coherent tunneling case. In the simple case of isotropic Fermi surfaces the in-plane projections of 3D Fermi momenta ${\bf Q}_\pm$ coincide with the Fermi momentum in 2D layer ${\bf k}_{2F}$.}
\label{fig-Fermi-surface-match}
\end{figure}
These mechanisms are mostly determined by the nature of the electron transfer between the two-dimensional (2D)
proximity system and the bulk SC. This transfer is strongly affected by both the mismatch of the band structures in the coupled subsystems and by disorder in the barrier between them. Without disorder and neglecting the band structure effects one arrives at the coherent tunneling model according to which the in-plane projection of the electron momentum is conserved in course of tunneling. The induced gap potential is determined by matching of the 2D Fermi surface with the in-plane projection of the 3D Fermi surface  (see Fig.~\ref{fig-Fermi-surface-match}). A generalization of the above model can include umklapp processes
accounting for the Bloch -- type single-electron wave functions in both subsystems. In the latter case, the momentum of tunneling electrons is conserved only up to certain vectors of the reciprocal lattices. One more limiting case is the so-called incoherent tunneling model which assumes a strong disorder in the tunneling barrier and allows for an arbitrary random change in the momenta of tunneling electrons.
The systematic analysis of these three tunneling models shows that the gap potential strongly depends on the degree of disorder as well as on the band structure effects.

Based on these models we consider several fundamental properties of the vortex matter in the systems with induced superconducting order.
First, the proximity induced superconducting gap $\Delta_{2D} $ is responsible for appearance of a new length scale in the vortex structure, the 2D coherence
length, $\xi_{2D} = \hbar v_{2F}/\Delta_{2D}$ or $\xi_{2D} = \sqrt{\hbar D_{2D}/\Delta_{2D}}$ for clean or dirty limits, respectively. Here $v_{2F}$
 and $D_{2D} $ are the Fermi velocity and diffusion constant in the 2D layer.
The energy gap $\Delta_{2D}$ depends on the tunneling rate $\Gamma$ \cite{KopninMelnikov11,AVolkov95,Fagas-etal-05,das-Sarma}; for example,
$\Delta_{2D}\approx \Gamma$ for $\Gamma \ll \Delta$.
  Since $\Delta_{2D}\ll\Delta$ the coherence length $\xi_{2D}$ usually is much longer than the coherence length in the bulk SC,
  $\xi_S = \hbar V_F/\Delta$ for clean or $\xi_S = \sqrt{\hbar D_S/\Delta} $ for dirty limit,
  where $\Delta$, $V_F$  and $D_S $ are the gap, the Fermi velocity and diffusion constant in the superconducting electrode.
  As a result, all the effects associated with overlapping of neighboring vortex cores as well as the normal QP scattering
  at the boundary of the 2D system become much more pronounced than in the primary superconducting electrode. There appears, e.g., an intriguing possibility to get a new type of vortex matter strongly bonded by the  intervortex QP tunneling even for magnetic fields well below the upper critical field of the bulk superconductor.

Second, hybridization of the localized QP states inside much larger induced vortex cores with the core states of primary vortices
 in the bulk electrode leads to peculiar structure of the subgap energy branches.
 For coherent tunneling, the electronic spectrum of a singly quantized vortex
consists of two
anomalous branches crossing zero of energy as functions of the impact parameter $b$. One
branch, $\epsilon_1 (b)$, qualitatively follows the
usual CdGM spectrum $\epsilon_0 (b)$ of the primary
vortex; it extends above the induced gap where it turns into a
scattering resonance. The other branch, $\epsilon_2
(b)$, lies below the induced gap and resembles the
CdGM spectrum for a vortex with a much larger core radius
$\sim\xi_{2D}$.
 Thus, the proximity induced vortex in a ballistic 2D layer has a
``multiple core'' structure characterized by the two length scales, $\xi_S$ and $\xi_{2D}$. Such a two-scale feature does not appear if the
proximity vortex states are induced by a primary vortex
pinned at a large-size hollow cylinder $r_0>\xi_S$, see Refs.~\cite{Rakhmanov11,Iosel_Feig}.

The spatial and energy
dependence  of the LDOS inside the multiple core reveals a rich
behavior which depends on many parameters and on the degree of disorder both inside the bulk electrode and inside the
2D layer, as well as by the barrier disorder. The barrier disorder
suppresses the influence of the primary CdGM spectral
branch and leads to broadening of
the lower anomalous branch $\epsilon_2 (b)$ due to the momentum uncertainty. Impurity
scattering in the bulk and/or inside the 2D layer causes
further smearing of the spectral characteristics of the core states which then approach the usual dirty-SC LDOS scaled with the corresponding coherence lengths $\xi_{2D}$.

And finally, both the nontrivial topological properties of the normal state wave functions and the induced pairing symmetry can affect the presence of the zero energy states in the QP spectrum of vortices. This phenomenon arises from the wave function symmetry under precession of the subgap QP trajectories inside the vortex core
through the corresponding change in the Bohr -- Sommerfeld quantization rule for the angular momentum.

The paper is organized as follows. In section \ref{sec-model} we introduce the basic model used further for the analysis of the induced superconductivity. The
derivation of self energies of 2D quasiclassical Eilenberger equations in a vortex state of the bulk superconductor is given in section~\ref{sec-ind-potentials}.
In section~\ref{sec-scale-sep-method} we discuss the method used for the calculation of the subgap state structure in the induced vortex core.
The main results are presented in sections~\ref{sec-ind-vort} and \ref{sec-ind-vort:subsec-disorder}.
In particular, section~\ref{sec-ind-vort} contains the results for the subgap spectrum and the local density of states in a induced vortex state of 2D layer.
In section~\ref{sec-discuss} we discuss implications of our analysis for induced vortex core states in graphene. We also discuss some further implications of a large value of the induced coherence length $\xi_{2D}$ for the spectral and spatial characteristics of various vortex configurations. Some details of our
calculations are given in Appendix.

\section{ Model.}\label{sec-model}

\begin{figure}[t]
\includegraphics[width=0.9\linewidth]{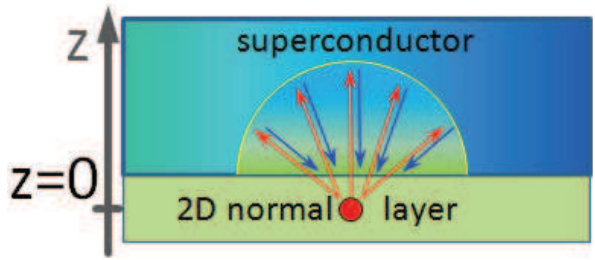}
\caption{(Color online) 2D normal metallic layer ($Z=0$) coupled to a bulk superconducting half-space $Z>0$ through a tunneling barrier. The electron waves depicted by red arrows tunnel from the source placed in 2D layer (red solid circle). If the energy is smaller than the superconducting gap, they do not penetrate deep into the bulk superconductor but undergo Andreev reflection to the hole waves (blue arrows) and return to the 2D layer.}
\label{fig-scheme}
\end{figure}

Consider a 2D normal metallic layer ($Z=0$) placed in a tunneling contact with  a bulk superconducting half-space $Z>0$ with a thin insulating barrier between them, see Fig.~\ref{fig-scheme}. The Hamiltonian of our system has the form
$\hat H = \hat H_{S}+\hat H_{2D}+ \hat H_{T}$,
where
\begin{multline}
\hat H_{S}= \int d^3R\left[\sum\limits_\sigma\hat\Psi^+_\sigma
({\bf X})\left(\hat\epsilon_{3D}-E_F\right)\hat\Psi_\sigma
({\bf X}) + \right.\\ \left.\Delta({\bf R})\hat\Psi^+_\uparrow
({\bf X})\hat\Psi^+_\downarrow
({\bf X})+\Delta^*({\bf R})\hat\Psi_\downarrow
({\bf X})\hat\Psi_\uparrow ({\bf X})\right]
\end{multline}
is the part describing the superconductor with the $s$-wave order parameter $\Delta({\bf R})$, $\hat \epsilon_{3D}$ is the kinetic energy operator, and
\begin{equation}
\hat H_{2D}= d\int d^2r\sum\limits_{\sigma}\hat
a^+_{\sigma}({\bf x})\left[\hat\epsilon_{2D}-E_F
\right]\hat
a_{\sigma}({\bf x})
\end{equation}
is the 2D layer Hamiltonian.
We introduce space-time variables ${\bf X}=({\bf R},\tau)$
and ${\bf x}=({\bf r},\tau)$ where ${\bf R}$ is a three-dimensional vector in the bulk superconducting
region while ${\bf r}$ is a two-dimensional vector in  the normal layer, respectively;
$\tau$ is an imaginary time variable in the standard Matsubara technique.
 The chemical potential $E_F$ is supposed to be equal in the subsystems.
 The single-particle Hamiltonian in the 2D layer $\hat\epsilon_{2D}$ includes the kinetic energy and, in general, the lattice potential corresponding to the crystal structure of the normal system. For simplicity we neglect the band structure of the bulk superconductor. This approximation should be valid for a wide class of heterostructures where the Fermi surface in the bulk SC is large compared with that in the 2D layer. We assume that tunneling is spin-independent and occurs locally in time and in space, i.e., from the point near the interface ${\bf R}=({\bf r}, Z=0)$ on the superconductor side into the point ${\bf r}$ in the layer and back with the amplitude $t({\bf r})$ that depends on the coordinate of the tunneling center on the interface. Since the tunneling
amplitude accounts for certain region of an atomic size in the vicinity of tunneling center, the wave function magnitude at $Z=0$ should be considered as an average value near the exact boundary of the superconducting region.
The tunneling amplitude is assumed small in the atomic scale. More detailed restrictions for the value of tunneling amplitude will be considered later.
The tunneling Hamiltonian has the form
\begin{multline}
\hat H_{T}=
d\sum\limits_{\sigma}\int\left[t({\bf r})\hat
\Psi^+_\sigma({\bf x})a_{\sigma}({\bf x})\right.\\ \left.+t^*({\bf r})\hat
a^+_{\sigma}({\bf x})\hat \Psi_\sigma({\bf x})\right]d^2 r
\end{multline}
where the wave functions in the superconductor are taken at the space-time point ${\bf x}$ at the interface $Z=0$.

The Matsubara Green
functions take the form:
\begin{subequations}\label{2D_Green_functions}
\begin{align}
\langle
T_\tau\hat a_{\alpha}({\bf x}_1)\hat
a^+_{\beta}({\bf x}_2)\rangle=\delta_{\alpha\beta}G({\bf x}_1,{\bf x}_2)
\ ,
\\
\langle
T_\tau\hat \Psi_{\alpha}({\bf X}_1)\hat
a^+_{\beta}({\bf x}_2)\rangle=\delta_{\alpha\beta}G_T({\bf X}_1,{\bf x}_2)
\ ,
\\
\langle
T_\tau\hat \Psi_{\alpha}({\bf X}_1)\hat
\Psi^+_{\beta}({\bf X}_2)\rangle=\delta_{\alpha\beta}G_S({\bf X}_1,{\bf X}_2)
\ ,
\end{align}
\end{subequations}
and
\begin{subequations}
\begin{align}
\langle
T_\tau\hat a_{\alpha}({\bf x}_1)\hat
a_{\beta}({\bf x}_2)\rangle=i\hat\sigma^{(y)}_{\alpha\beta}
F({\bf x}_1,{\bf x}_2) \ ,
\\
\langle
T_\tau\Psi_{\alpha}({\bf X}_1)\hat
a_{\beta}({\bf x}_2)\rangle=i\hat\sigma^{(y)}_{\alpha\beta}
F_T({\bf X}_1,{\bf x}_2) \ ,
\\
\langle
T_\tau\hat \Psi_{\alpha}({\bf X}_1)\hat
\Psi_{\beta}({\bf X}_2)\rangle=i\hat\sigma^{(y)}_{\alpha\beta}
F_S({\bf X}_1,{\bf X}_2) \ ,
\end{align}
\end{subequations}
etc.
Equations for the Green functions can be more conveniently written in the frequency representation $\omega_n=(2n+1)\pi T$. We denote $\tau=\tau_1-\tau_2$ and write
$$
G({\bf r}_1,{\bf r}_2)=\int_0^{\hbar/T} G({\bf r}_1,{\bf r}_2;\tau)e^{i\omega_n\tau/\hbar} d\tau \ ,
$$
skipping for simplicity the subscript. We introduce also the Nambu matrices for Hamiltonian and Green functions
$$\check H_S=\begin{pmatrix}
\hat \epsilon_{3D}-E_F & -\Delta({\bf R})\\
\Delta^*({\bf R}) & \hat \epsilon_{3D}-E_F\\
\end{pmatrix}, \quad
\check G = \begin{pmatrix}
G & F \\
-F^+ & \bar{G}
\end{pmatrix} \ ,
$$
and denote the inverse operators
\begin{eqnarray*}
\check G_S^{-1}({\bf R})&=&
-i\check \tau_3\omega_n +\check H_S, \\
\check G_{2D}^{-1}({\bf r})&=&
-i\check \tau_3\omega_n +\check \tau_0\otimes\left[\hat\epsilon_{2D}-E_F
\right] \ ,
\end{eqnarray*}
in the superconductor and 2D layer, respectively.

Equations for the mixed Green functions $\check G_T({\bf R}_1,{\bf
r}_2) $ can be written in the form
\begin{equation*}
\check G_S^{-1}({\bf R}_1)\check G_T({\bf R}_1,{\bf r}_2)+d
\check t({\bf R}_{1\perp}) \check G({\bf R}_{1\perp},{\bf r}_2)\delta(Z_1) =0
\end{equation*}
where $Z_1\geq0$, ${\bf R}_1 = ({\bf R}_{1\perp},Z_1)$ and
\begin{equation*}
\check t({\bf r})= \left(\begin{array}{lr} t({\bf r}) & 0 \\
0 & t^*({\bf r})\end{array}\right)\ .
\end{equation*}
Neglecting the back-action of
a thin 2D layer on the superconductor, we assume that the superconducting Green function $\check G_S({\bf R}_1,{\bf R}_2)$ is a non-interacting function that satisfies
\begin{equation}\label{G_S_eq}
\check G_S^{-1}({\bf R}_1)\check G_S({\bf R}_1,{\bf R}_2) =\check
1 \hbar\delta({\bf R}_1-{\bf R}_2)
\end{equation}
in the range $Z_{1,2}>0$. The boundary conditions for $\check G_S({\bf R}_1,{\bf R}_2)$ at $Z=0$ depend on
the particular interface in the absence of tunneling. This gives
\begin{equation}
\check G_T({\bf R}_1,{\bf r}_2)
=-\frac{d}{\hbar} \int \check G_S({\bf R}_1,{\bf r}^\prime) \check t({\bf r}^\prime) \check G({\bf r}^\prime ,{\bf
r}_2)\, d^2 r^\prime  \label{G-across}
\end{equation}
%

Equations for the Green functions in the layer can be written as
\begin{multline*}
\check G_{2D}^{-1}({\bf r}_1)\check G({\bf r}_1,{\bf r}_2)
+ \check t^* ({\bf r}_1) \check
G_T({\bf r}_1,{\bf r}_2) \\
 =\check 1 d^{-1} \hbar\delta
({\bf r}_1 -{\bf r}_2)
\end{multline*}

Using Eq.~\eqref{G-across} we find
\begin{multline}\label{eqG-coord}
\check G_{2D}^{-1}({\bf r}_1)\check G({\bf r}_1,{\bf r}_2)
-\int \check \Sigma _T({\bf r}_1,{\bf r}^\prime )\check G({\bf r}^\prime ,{\bf r}_2)\, d^2 r^\prime
\\
=\check 1 \hbar d^{-1} \delta ({\bf r}_1 -{\bf
r}_2) \ ,
\end{multline}
where
\begin{equation}\label{selfen-gen}
\check \Sigma _T({\bf r}_1,{\bf r}^\prime )=\left(\begin{array}{cc} \Sigma_1 & \Sigma_2\\
- \Sigma_2^\dagger & \bar \Sigma_1\end{array}\right) =\frac{d}{\hbar} \check
t^* ({\bf r}_1) \check G_S^{0}({\bf r}_1,{\bf r}^\prime)\check t({\bf r}^\prime ).
\end{equation}

One can introduce the momentum representation of the Green function \cite{Kopnin-book}
\begin{equation}\label{p-represent}
\check G_S({\bf R}_1,{\bf R}_2)=\int\frac{d^3Q_1}{(2\pi)^3}\frac{d^3Q_2}{(2\pi)^3}\check G_S({\bf Q}_1,{\bf Q}_2)e^{i {\bf Q}_1{\bf R}_1-i {\bf Q}_2{\bf R}_2} \ .
\end{equation}
and the tunneling coefficients: $\check t({\bf r})=\int\frac{d^2 q}{(2\pi)^2}\check t({\bf q})e^{i {\bf q}{\bf r}}$.
The Fourier representation for the Green functions in the 2D layer is
\begin{equation}
\check G ({\bf r}_1,{\bf r}_2)=\int\frac{d^2q_1}{(2\pi)^2}\frac{d^2 q_2}{(2\pi)^2}\check G ({\bf q}_1,{\bf q}_2)e^{i {\bf q}_1{\bf r}_1-i {\bf q}_2{\bf r}_2} \ .
\end{equation}

\subsection{Tunneling with umklapp processes.}

The crystal structure of the 2D layer accounts for an atomic-scale periodic potential
in Eq.~\eqref{eqG-coord} which mixes the
Fourier harmonics with the momenta shifted by the reciprocal lattice vectors ${\bf b}$.
Using the Bloch functions
$$\psi_m({\bf k, r})=\sum\limits_{{\bf b} } e^{i({\bf k+b}){\bf r}}u_{m {\bf k+b}}$$
 diagonalizing the single-particle
energy operator inside the layer
$$\epsilon_{2D}({\bf r})\psi_m({\bf k, r})=\epsilon_m(k)\psi_m({\bf k, r})$$
one can conveniently introduce
 the field operators $\hat a_{\alpha,m,{\bf k}}$
$$\hat a_\alpha({\bf r})=\sum\limits_{m}\int\frac{d^2 k}{(2\pi)^2}\hat a_{\alpha,m,{\bf k}} \psi_m({\bf k, r}) \ .$$
The index $m$ enumerates the energy bands.

Introducing
the corresponding Green functions
\begin{subequations}\label{2D_Green_function_band_repr}
\begin{align}
\langle
T_\tau\hat a_{\alpha, m_1, {\bf k}_1}\hat
a^+_{\beta, m_2, {\bf k}_2}\rangle=\delta_{\alpha\beta}G_{m_1,m_2}({\bf k}_1,{\bf k}_2)
\ ,\\
\langle
T_\tau\hat a_{\alpha, m_1, {\bf k}_1}\hat
a_{\beta, m_2, -{\bf k}_2}\rangle=i\hat\sigma^{(y)}_{\alpha\beta}
F_{m_1,m_2}({\bf k}_1,{\bf k}_2) \ ,
\end{align}
\end{subequations}
one can diagonalize the operator $\check G_{2D}^{-1}$ in Eq.~\eqref{eqG-coord}
in the Bloch representation,
\begin{equation}
\check G_{2D,m}^{-1}({\bf k})=-i\hbar \check\tau_3\omega_n + \begin{pmatrix} \epsilon_{m}({\bf k})-E_F & 0\\ 0 & \epsilon_{m}(-{\bf k})-E_F\end{pmatrix}\ .
\end{equation}
We assume in what follows that the amplitude $\Delta_{ind}$ of the induced superconducting gap $\Delta_{2D}$ is small compared to the interband distance $\epsilon_m - \epsilon_{m'}$ and neglect the interband scattering. Hereafter
we omit the subscripts $m$.
At the same time, the transformation from the momentum to the quasimomentum  representation results in the mixing of Fourier harmonics in the self energy in Eq. \eqref{eqG-coord}. Finally, Eq.\ \eqref{eqG-coord} for the Green functions \eqref{2D_Green_function_band_repr} takes the form:
\begin{multline}\label{G_2Dn_eq}
\check G_{2D}^{-1}({\bf k}_1)\check G({\bf k}_1,{\bf k}_2)-
\int\check \Sigma_{T}({\bf k}_1,{\bf k}')\check G({\bf k}',{\bf k}_2)d^2 k'\\={\hbar}\check 1 \delta({\bf k}_1-{\bf k}_2) \ ,
\end{multline}
with
$$\check \Sigma_{T}({\bf k}_1,{\bf k}')=\frac{d}{\hbar}\int \check t_b^+({\bf k}_1,{\bf Q}_\perp)\check G_{S}^0({\bf Q},{\bf Q}')\check t_{b}({\bf Q}'_\perp,{\bf k}')d^3 Q d^3 Q' \ ,$$
\begin{equation}\label{T_n}
\check t_b({\bf Q}, {\bf k})=
\sum\limits_{{\bf b}} u_{{\bf k+b}} \check t({\bf Q_\perp-k-b})
\end{equation}
and $\check t_b^+({\bf k, Q}_\perp)=\check t_b^*({\bf Q_\perp, k})$. Here ${\bf Q}=({\bf Q}_\perp,Q_z)$.
The above expression for the tunneling coefficients $t_b$ describes in fact the umklapp processes caused by the periodic crystal
potential in the 2D layer.

\subsection{Coherent tunneling}

The simplest model of tunneling assumes that the in-plane momentum projection of electrons is
conserved during the tunneling process: $\check t({\bf Q_\perp-k})=\check t\cdot \delta({\bf Q_\perp-k})$.
This is equivalent to the assumption that the tunneling amplitude $t({\bf r})$ is independent of the coordinate
along the SC/2D interface.
Of course, the quasimomentum conservation is not exact in the presence of energy bands since
the tunneling mixes the quasimomentum values which differ by a reciprocal lattice vector:
 $\check t_b({\bf Q}, {\bf k})=
\check t\sum_{{\bf b}} u_{{\bf k+b}} \delta({\bf Q_\perp-k-b})$.
Neglecting umklapp processes for simplicity we find from Eq.~\eqref{G_2Dn_eq}
\begin{equation*}
\check \Sigma _T({\bf k}_1,{\bf k}^\prime)= \frac{dt^2}{\hbar}\int \check G_S^{0}
({\bf k}_1,Q_z;{\bf k}^\prime, Q_z^\prime)\frac{dQ_z\,
dQ_z^\prime}{(2\pi)^2} \ .
\end{equation*}

From now on we will use the quasiclassical approximation for the Green functions. In order to derive the Eilenberger equations in the 2D layer we follow the standard procedure described, e.g., in Ref.~\onlinecite{Kopnin-book}.
First of all we introduce the average ${\bf k} =({\bf k}_1+{\bf k}_2)/2$, $Q_z=(Q_{1z}+Q_{2z})/2$ and relative ${\bf k}_- ={\bf k}_1-{\bf k}_2$, $q_z=Q_{1z}-Q_{2z}$ momenta and denote $\check G({\bf k}_1,{\bf k}_2)=\check {\mathcal G}({\bf k},{\bf k}_-)$, $\check G_S({\bf k}_1, Q_{1z};{\bf k}_2,Q_{2z})=\check {\mathcal G}_S({\bf k},Q_z;{\bf k}_-, q_z)$.
Next we apply the operator $\check G_{2D}^{-1}$ to the Green function $\check {\mathcal G}({\bf k},{\bf k}_-)$ from the right
and subtract this equation from Eq.~\eqref{G_2Dn_eq}. We now transform to the quasiclassical Green functions by integrating the resulting equation over $d\xi_2$ where $\xi_2=\epsilon_{2D}({\bf k}) - E_F$.
The Green functions are to be taken in the vicinity of the Fermi surface.
Therefore, in the mixed momentum-coordinate representation,
\begin{eqnarray*}
\check {\mathcal G}({\bf k},{\bf r})&=&\int \check {\mathcal G}({\bf k},{\bf k}_-)e^{i{\bf k}_-{\bf r}}\, \frac{d^2 k_-}{(2\pi)^2}\\
\check {\mathcal G}_S({\bf k},Q_z;{\bf r},Z)&=&\int \check {\mathcal G}_S({\bf k},Q_z;{\bf k}_-, q_z)e^{i{\bf k}_-{\bf r}+iq_z Z}\, \frac{d^2 k_-dQ_z}{(2\pi)^3}
\end{eqnarray*}
we can put
\begin{eqnarray*}
\check {\mathcal G}_S ({\bf k},Q_z;{\bf r}, Z)&=
&\check g_S ({\bf k},Q_{z};{\bf r}, Z) \pi i \delta_{\Delta}(\xi_3) \ ,\\
\check {\mathcal G} ({\bf k} , {\bf r})&=&\check g({\bf k}
,{\bf r}) \pi i \delta_{\Delta}(\xi_{2})
\end{eqnarray*}
Here the standard quasiclassical Green functions are
\begin{eqnarray}
\check g ({\bf
k}_{2F},{\bf r})&=&\frac{1}{\pi i}\int {d\xi_{2}} \check {\mathcal G} ({\bf
k},{\bf r}) \ ,\\
\check g_S ({\bf
K}_F,{\bf R})&=&\frac{1}{\pi i}\int {d\xi_{3}} \check {\mathcal G}_{S} ({\bf
Q},{\bf R}) \ .
\end{eqnarray}
$\xi_3=\epsilon_{S}({\bf Q})-E_F$ is the normal QP
spectrum in the 3D half-space, and
$\delta_{\Delta}(\xi_{2,3})$ is a delta function broadened at the
gap energy scale $\Delta$.

At the next step of derivation we note that, in the mixed representation, the term
\begin{equation*}
\int \frac{d\xi_{2}}{\pi i} \int \check \Sigma _T({\bf k}_1,{\bf
k}^\prime
)\check G({\bf k}^\prime ,{\bf k}_2)\frac{d^2 k^\prime}{(2\pi)^2}
\end{equation*}
in the equation for the Green function becomes
\begin{multline*}
\frac{\pi i dt^2}{\hbar} \int  d\xi_2 \int \frac{dQ_z}{2\pi} \check g_S ({\bf k},Q_{z};{\bf r}, 0)
\check g({\bf k},{\bf r}) \delta_{\Delta}(\xi_3) \delta_{\Delta}(\xi_{2})\\
=\frac{\pi i dt^2}{\hbar}  \int \frac{dQ_z}{2\pi} \check g_S ({\bf Q};{\bf r}, 0)
\check g({\bf k}_{2F},{\bf r})
\times \delta_{\Delta}[\epsilon_{3D}({\bf Q})-E_F]
\end{multline*}
where ${\bf Q}=({\bf k}_{2F},Q_z)$ has the in-plane projection coinciding with the 2D Fermi momentum ${\bf k}_{2F}$.

Finally, we obtain the quasiclassical Eilenberger equation for retarded (advanced) Green functions
\begin{multline} 
-i\hbar{\bf v}_{2F} {\bm \nabla}\check g({\bf k}_{2F},{\bf r}) -
\epsilon \left[ \check \tau_3 \check g({\bf k}_{2F} ,{\bf
r})-\check g({\bf k}_{2F} ,{\bf r})\check \tau_3
\right] \\
-\left[\check \Sigma_T \check g({\bf k}_{2F},{\bf r}) - \check
g({\bf k}_{2F},{\bf r})\check \Sigma_T \right]=0 \
,\label{eq-Eliashb-st}
\end{multline}
where $\hbar{\bf v}_{2F}=\partial\epsilon_{2D}({\bf
k})/\partial{\bf k}$ is the 2D layer Fermi velocity.

For isotropic Fermi surfaces in both the
superconductor $\epsilon_{3D}({\bf Q})=\hbar^2Q^2/2m$ and the 2D layer $\epsilon_m({\bf k})=\hbar^2 k^2/2 m_{2D}$, the self energy takes the form
\begin{equation}
\check \Sigma_T ({\bf k}_{2F},{\bf r}) = \frac{i\Gamma}{2} \left[
\check g_S ({\bf Q}_{+};{\bf r}, 0) + \check g_S ({\bf Q}_{-};{\bf r}, 0)\right], \label{selfen-coh}
\end{equation}
with the tunneling rate
$$
\Gamma =  dt^2 \int_0^\infty
\delta_{\Delta}\left[\epsilon_S({\bf k}_{2F},Q_z)-E_F\right] dQ_z\
.
$$
The 3D momentum ${\bf Q}_\pm = ({\bf k}_{2F}, \pm Q_{3z})$ lies on
the Fermi surface of the bulk SC, $k^2_{2F}+Q_{3z}^2=K_F^2$.
Provided the 2D Fermi surface is smaller than the extremal cross
section of the 3D Fermi surface, i.e., $k_{2F}<K_F$ the expression
for the tunneling rate reads: ${\Gamma=d m t^2/Q_{3z}}$. For large
2D Fermi surfaces $k_{2F}>K_F$ the self -- energy term vanishes, and
the coherent tunneling is impossible. The case of  momenta
$k_{2F}\simeq K_F$ deserves special consideration which should
take account of a finite delta function width:
 $\Gamma\sim d t^2 (m/\Delta)^{1/2}$.

The umklapp processes should, of course, modify the self -- energy part resulting in additional
contributions:
\begin{equation}
\check \Sigma_T ({\bf k}_{2F},{\bf r}) = \sum_{\bf b} |u_{{\bf k}_{2F}+{\bf b}}|^2 \check \Sigma_T^{(0)} ({\bf k}_{2F}+{\bf b},{\bf r}), \label{selfen-umklapp}
\end{equation}
where
$\check \Sigma_T^{(0)} ({\bf k}_{2F},{\bf r})$ is given by the Eq.\eqref{selfen-coh}.

\subsection{Incoherent tunneling}

The coherent tunneling model in many cases oversimplifies the realistic experimental situation. The momentum conservation is violated, for example, by the presence of disorder at the interface. Here we consider an opposite limit of strong disorder, which is sometimes called the incoherent tunneling model.
This model assumes a random tunneling process of electrons through the barrier in a way similar to the standard theory of dirty metals within the Born approximation \cite{AGD}. We assume that
the ensemble average of tunneling amplitudes is
\begin{equation}
\overline{t({\bf r}_1) t({\bf r}_2)} = t^2 s_a \delta({\bf
r}_1-{\bf r}_2 ) \ ,
\end{equation}
where $s_a$ is the correlated area of the order of atomic
scale.
Following the standard diagrammatic procedure we expand the solution for the ensemble averaged Green function in a series in
the scattering field and split the multiple correlators of the $t({\bf r})$ values in a product of the above pair correlators.
Finally, after averaging the self energy \eqref{selfen-gen} becomes:
\begin{multline}
\check \Sigma _T({\bf r}_1,{\bf r}_2)=t^2 ds_a\check G_S({\bf r}_1,{\bf r}_1;0)\delta({\bf r}_1-{\bf r}_2 )\\
= t^2 ds_ai\pi \nu_3(0) \left<\check g_S({\bf Q};{\bf
r},0)\right>\delta({\bf r}_1-{\bf r}_2 ) \ .
\end{multline}
Here $\nu_3(0)$ is the normal density of states in the bulk material.
Angular brackets denote averaging over
three-dimensional momentum directions. Within the quasiclassical
approach, the resulting self energy to be used in the Eilenberger equation (\ref{eq-Eliashb-st}) is given by
\begin{equation}
\check \Sigma _T({\bf r})=i\Gamma \left<\check g_S({\bf Q};{\bf
r},0)\right>\ . \label{selfen}
\end{equation}
where the tunneling rate is $\Gamma =\pi
\nu_3(0)ds_at^2 $. This approximation coincides with
that used in Ref.~\onlinecite{KopninMelnikov11}.
The tunneling rate $\Gamma \sim t^2/E_F$ can be expressed
\cite{KopninMelnikov11}
 in terms of the normal-state tunnel conductance $G=1/RS$ per unit contact area,
$
\Gamma=G/(4\pi G_0\nu_2) \sim E_F R_0 /R \ ,
$
with the conductance quantum $G_0=e^2/\pi \hbar$ and
the normal 2D density of states (DOS) $\nu_2=m_{2D}/2\pi \hbar^2$.
Therefore $ \Gamma/E_F\ll 1$ if the total tunnel resistance $R$
is much larger than the  Sharvin resistance $R_0=(NG_0)^{-1} $
for an ideal $N$-mode contact with the contact area $S$. 
 Nevertheless, there is a room for the condition
 $\Gamma \sim \Delta$ to be fulfilled even for the large contact resistance $R\gg R_0$.

\subsection{Adiabatic approximation. Range of validity.}\label{sec-model:subsec-phenom}

The above microscopic analysis allows us to comment on the simplest phenomenological
model which is often used for description of the proximity induced superconductivity, see for example, \cite{been2,Oreg10,Iosel_Feig,Perfetto13}.
Within this model, the Bogoliubov -- de Gennes equations inside the proximity superconductor include a
phenomenological gap function which is postulated to be proportional to the gap function $\Delta$ inside the superconducting electrode. Our approach shows that this is generally not the case. The true equation (\ref{eq-Eliashb-st}) includes self-energies which are complicated functions of energy, coordinates, and momentum. In fact, the effective gap function resembles that in a usual superconductor only if the bulk SC is homogeneous in space. In this case, the quasiclassical Green function is
\begin{equation*}
\check g^{R(A)}_\epsilon =\pm \frac{1}{\sqrt{\epsilon^2 -|\Delta|^2}}\left(\begin{array}{cc} \epsilon & \Delta \\
-\Delta^* & -\epsilon\end{array}\right)
\end{equation*}
In this case the self energy is $\check \Sigma _T = i\Gamma \check g_S$, for both coherent and incoherent
tunneling models. This expression also holds if the superconducting gap is a slow function of coordinates on distances of the order of $\xi_S$. For $|\epsilon|<|\Delta|$ the self-energy has the form
\begin{equation}
 \check\Sigma_T({\bf r}) =\frac{\Gamma}{\sqrt{|\Delta({\bf r})|^2-\epsilon^2}}\begin{pmatrix}
\epsilon & \Delta({\bf r})\\ -\Delta^*({\bf r})&-\epsilon
 \end{pmatrix} \ . \label{sigma-adiabat}
 \end{equation}
Only for low-transparency tunnel contact, $\Gamma \ll \Delta$, this self energy is nearly off-diagonal on the scale $\epsilon \sim \Gamma$ and can be regarded as an energy-independent effective gap function
\begin{eqnarray}\label{fg-asimpt_gen}
\check \Sigma _T\simeq
i\Gamma\check\tau_2e^{i\check\tau_3\phi} \ ,
\end{eqnarray}
where $\phi$ is the phase of the superconducting order parameter.
Note that the resulting induced gap does not at all depend on the gap magnitude $|\Delta|$ in the bulk. If the transparency is finite, the electronic spectrum in the induced superconductor has a gap $\Delta_{2D}$ which is determined by the condition \cite{KopninMelnikov11,McMillan}
\begin{equation} \label{induced_gap}
(\epsilon + \Sigma_1)^2 -\Sigma_2^2 =0\ , \; \epsilon =\Delta_{2D} \ .
\end{equation}
Of course, the adiabatic approximation also breaks down if the order parameter $\Delta$ varies as a function of coordinates at distances of the order of coherence length in the superconducting electrode, when the self-energies are no longer determined by Eq.~\eqref{sigma-adiabat}.

\section{Vortex potentials and Green functions for clean systems} \label{sec-ind-potentials}

The quasiclassical Green functions in the 2D
layer satisfy the Eilenberger equations \eqref{eq-Eliashb-st}. In components,
\begin{subequations}\label{Elen}
\begin{align}
-i\hbar{\bf v}_{2F}  {\bm \nabla}  f -2\left(\epsilon +\Sigma_1 \right)\! f +2 \Sigma_2 g =0 ,\; \label{El2} \\
i\hbar{\bf v}_{2F}  {\bm \nabla}  f^{\dagger } -2\left(\epsilon +\Sigma_1  \right)\! f^{\dagger} +2\Sigma_2^{\dagger} g =0 ,\;
\label{El3}\\
-i\hbar{\bf v}_{2F}  {\bm \nabla}  g+ \Sigma _2 f ^{\dagger} - \Sigma^\dagger _2 f =0 .\; \label{El1}
\end{align}
\end{subequations}
and the normalization condition $g ^2- f  f^{\dagger} =1$ with the
self energies Eqs.~(\ref{selfen-coh}) or  (\ref{selfen})
as effective potentials.

\begin{figure}[t]
\includegraphics[width=0.75\linewidth]{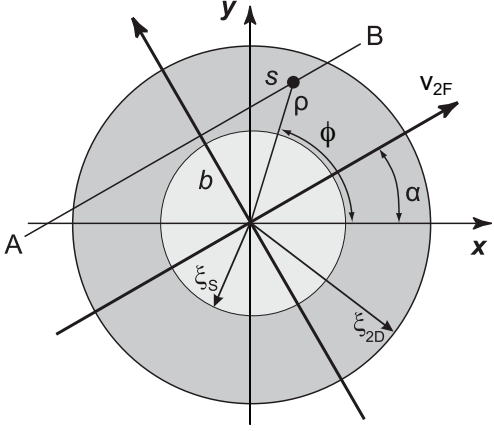}
\caption{(Color online) 
  The coordinate frame near the
multiple vortex core. Primary (induced) core is shown by the white (gray) circle. The QP trajectory
with an impact parameter $b$ (line AB) passes through the point
$(\rho ,\phi )$ shown by the black dot.}
\label{fig-2Dvortex-coords}
\end{figure}

In this and the following Section we consider the case of isotropic Fermi surfaces. Modifications due to the anisotropy of the spectrum are discussed in Section \ref{sec-anisotropic}.
QPs in clean systems are conveniently described by the coordinates along their trajectories (see Fig.~\ref{fig-2Dvortex-coords}). A quasiclassical trajectory is parameterized by its angle $\alpha$ with the $x$ axis, the impact parameter $b=\rho \sin (\phi -\alpha )$ and the coordinate $s=\rho
\cos (\phi -\alpha )$ along the trajectory. We introduce the
symmetric and antisymmetric parts of the Green functions as it was done in
\cite{KramerPesch, Kopnin-book}:
\begin{subequations}\label{f-param}
\begin{align}
f=-\left[\zeta (s)+i\theta (s)\right]\exp (i\alpha)\\
f^{\dagger }=\left[\zeta (s)-i\theta (s)\right]\exp (-i\alpha ),
\end{align}
\end{subequations}
where $\zeta (s)=\zeta (-s)$, and $\theta (s)=-\theta (-s)$. The
normalization condition requires $g^{2}+\theta ^{2}+\zeta ^{2}=1$.
Eilenberger equations \eqref{Elen} can be rewritten in the form
\begin{subequations}\label{eq-zeta-theta-g-2}
\begin{align}
\hbar v_{2F} \frac{d\zeta}{ds} +2\left(\epsilon +\Sigma_1\right) \theta  -2ig\Sigma_R &=0  ,\quad \label{eq-zeta-2}\\
\hbar v_{2F} \frac{d\theta }{ds} -2\left(\epsilon + \Sigma_1\right) \zeta  -2ig \Sigma_I &=0 , \quad \label{eq-theta-2}\\
\hbar v_{2F}\frac{dg}{ds} +2i\zeta\Sigma_R+2i\theta\Sigma_I &=0 , \quad  \label{eq-g-2}
\end{align}
\end{subequations}
where
\begin{subequations}\label{Sigma_RI}
\begin{align}
2\Sigma_R=\left( \Sigma_2  e^{-i\alpha} +  \Sigma_2^{\dagger}e^{i\alpha} \right) ,\quad\\
2i\Sigma_I=\left( \Sigma_2  e^{-i\alpha} -  \Sigma_2^{\dagger}e^{i\alpha} \right) .\quad
\end{align}
\end{subequations}

In the present paper we consider the limit of low tunneling rate $\Gamma \ll \Delta$
which leads to a small induced gap \cite{KopninMelnikov11} $\Delta_{2D}=\Gamma$ and long coherence length $\xi_{2D}\gg \xi_{S}$.
We consider an isolated vortex line oriented along the $Z$ axis perpendicular to
the SC/2D interface and choose the gap function inside the bulk
SC in the form ${\Delta =\Delta_0(\rho) e^{i\phi}}$,
where $(\rho , \phi)$ are the cylindrical coordinates;  $\Delta_0(\rho)$ approaches the bulk value
$\Delta_\infty$ far from the vortex core.  The
self energies in the 2D layer,
Eqs.~(\ref{selfen-coh}) or (\ref{selfen}), have parts with sharp peaks localized at small
distances $\rho\sim\xi_S$ and the adiabatic long-distance ``vortex potential'' tail
$\Delta_{2D}\sim\Gamma  e^{i\phi}$ at $\rho \gg \xi_S$ according to Eq.~\eqref{fg-asimpt_gen}.

In the case of clean bulk SC we use the condition of specular reflection at the interface. This can be applied for both coherent and incoherent tunneling models since any possible disorder in tunneling affects only a tiny fraction of bulk electrons whose vast majority reflects without tunneling.
For specular reflection, one can
use the bulk quasiclassical Green functions obtained for an infinite space. For energies $\epsilon\ll\Delta_\infty$,  the
self energy, Eq.~\eqref{fg-asimpt_gen} for long distances ($\rho\gg\xi_S$) is
independent of the particular tunneling model and of the disorder
in the bulk SC: $\Sigma_1 \approx 0$, $\Sigma_2\approx \Gamma e^{i\phi}$, i.e., $\Sigma_R\approx \Gamma
s/\rho$, $\Sigma_I\approx\Gamma b/\rho$.
However, the induced vortex potentials
close to the primary  vortex core are very sensitive
to the impurity concentration and momentum exchange
during the tunneling process.

For clean bulk SC, the Green function can be parameterized
similar to \eqref{f-param} with $f\to f_S$, $\zeta\to\zeta_S$, and $\theta\to\theta_S$.
The Eilenberger equations have the form of Eqs.~\eqref{eq-zeta-theta-g-2}
with $v_{2F} \to v_{\parallel}=V_F \cos\chi_p$ where $\chi_p$ is the polar angle of the momentum, while $\Sigma_1=0$, $\Sigma_2\to \Delta=\Delta_0(\rho)e^{i\phi}$,
 and $\Sigma_2^\dagger \to \Delta^*$.
For energies  $\epsilon\ll \Delta_\infty$ and distances $s$ of the order or less than the core size, the functions $g_S$ and $f_S$
are given in Refs.~\cite{KramerPesch,Kopnin-book}.
\begin{eqnarray}
\zeta_S  =\frac{\hbar v_{\parallel }e^{-K}}{2\Lambda\left[ \epsilon -\epsilon
_{0}\pm i\delta \right]} , \; \theta_S  = \frac{2}{\hbar v_{\parallel
}}\int_{0}^{s}(\epsilon - \frac{b\Delta_0}{\rho^\prime}
)\zeta_S ds^\prime , \qquad  \label{zeta}\\
\epsilon _{0}(b) =b \Lambda^{-1}\int_{0}^{\infty }[\Delta _{0}/\rho ]%
e^{-K(s)}\,ds  ,\qquad  \label{Eb}\\
\Lambda =\int_{0}^{\infty }e^{-K(s)}\,ds \ ;\quad K(s)=
\frac{2}{\hbar v_{\parallel}}\int_{|b|}^{\rho}\Delta
_{0}(\rho^\prime) \,d\rho^{\prime }
 \ . \qquad  \label{K}
\end{eqnarray}
For larger distances, $s\gg \xi_S$, the
function $\zeta_S$ assumes its asymptotic expression
$\zeta^{R(A)}_S=-b/\rho$ corresponding to the boundary conditions
Eq.~\eqref{fg-asimpt_gen}.

\subsection{Vortex potentials for coherent tunneling}

The vortex potentials induced in the 2D layer
crucially depend on the tunneling mechanism.
For example, within the coherent tunneling model we get
\begin{equation*}
\Sigma_1 =
i\Gamma g_S({\bf Q},{\bf r}), \Sigma_2 = i\Gamma f_S({\bf Q},{\bf r})
\end{equation*}
in terms of the infinite-space Green functions, since $\check g_S(+Q_{3z})=\check g_S(-Q_{3z})$ for specular reflection.
For energies  $\epsilon\ll \Delta_\infty$ and distances $s$ of the order or less than the core size $\xi_S$, we find from Eq. (\ref{f-param})
\begin{subequations} \label{sigma-coherent}
\begin{align}
\Sigma_1 &=-
\Gamma \zeta_S, \;\\
 \Sigma_2 &= \Gamma [\theta_S-i\zeta_S ]e^{i\alpha}, \;\\
 \Sigma_2^\dagger &= \Gamma [\theta_S+i\zeta_S ]e^{-i\alpha} \ ,
\end{align}
\end{subequations}
where $\zeta_S$ and $\theta_S$ are given by Eqs. (\ref{zeta})--(\ref{K}).

\subsection{Vortex potentials for incoherent tunneling}

For incoherent tunneling,  we find $\Sigma_1 =
i\Gamma \left<g_S\right>$, $\Sigma_2 = i\Gamma \left<f_S\right>$ where averaging over the 3D momentum direction is equivalent to the ensemble averaging.
To calculate the angular average
one can separate the Green functions into the principal-value part and the
delta-functional contribution. For example,
\begin{equation}
g_S^{R(A)}=i\zeta_S^{R(A)} = \wp \frac{i \hbar v_{\parallel
}e^{-K}}{2\Lambda \left[ \epsilon -\epsilon _{0}\right] }\pm  \frac{\pi
\hbar v_{\parallel }e^{-K}}{2\Lambda }\delta(  \epsilon -\epsilon _{0}) \ .
\label{Gfunct/vortcore}
\end{equation}

Performing averaging over the polar $\chi_p$ and azimuthal
$\alpha$  angles we take into account the symmetry of the
functions under the $s$-inversion transformation. As a result, we obtain
\begin{eqnarray}
\Sigma_1&=& -\Gamma\left<\zeta_S(s)\right>\ , \label{Sigma_1_incoh}\\
\Sigma_2e^{-i\phi}&=& \Sigma_2^{\dagger} e^{i\phi}= \Sigma_{ad}+\Sigma_2^{loc}
\ ,\label{Sigma_I_incoh} \\
\Sigma_{ad}(\rho)&=&
 \wp  \left<\, {\Gamma I(s){\rm sign} (s)}/{2\Lambda \left[ \epsilon -\epsilon _{0}\right]
}\right> \ .\label{Phi}
\end{eqnarray}
We put here
$$I(s)=2\int_{0}^{s}(\epsilon -\Delta _{0}b/\rho
)e^{-K(s^{\prime })}\,ds^{\prime } \ .$$
The off-diagonal components of induced potential are split into
the localized and the long-range parts, $\Sigma_2^{loc}$ and $\Sigma_{ad}$, respectively.  The long-range function $\Sigma_{ad}$ can be regarded as an adiabatic induced superconducting
gap, $\Sigma_{ad} \to \Gamma$ for $\rho \gg \xi_S$ and $\Sigma_{ad} \to 0$ for $\rho \to 0$.  Averaging over
the azimuthal trajectory angle $\alpha$ we find:
\begin{eqnarray*}
\Re \Sigma_2^{loc}&=&\Gamma\left<\frac{\hbar v_\parallel e^{-K}}{2\Lambda \Omega\rho }
\left[1-\Re\frac{|\epsilon|}{\sqrt{\epsilon^2 -\Omega^2\rho^2}}
\right]\right>_z \ ,\\
\Im \Sigma_2^{loc}&=&
\pm\Gamma\left< \Re\frac{\epsilon \hbar v_\parallel e^{-K}}
{2\Lambda \Omega \rho \sqrt{\Omega^2\rho^2 -\epsilon^2}}
\right>_z \ ,\\
\Re \Sigma_1
&=&-{\rm sign}(\epsilon) \Gamma \left< \Re\frac{\hbar v_\parallel e^{-K}}
{2\Lambda \sqrt{\epsilon^2-\Omega^2\rho^2}}
\right>_z  \ ,\\
\Im \Sigma_1
&=& \pm\Gamma\left<\Re\frac{\hbar v_\parallel e^{-K}}
{2\Lambda \sqrt{\Omega^2\rho^2 -\epsilon^2}}
\right>_z \ .
\end{eqnarray*}
Here the upper (lower) sign corresponds to a retarded (advanced)
self energy term, $\Omega = d\epsilon_0/db$,
and we use the notation
\begin{equation*}
\left< \ldots\right>_z = \frac{1}{2} \int_0^\pi \left(\ldots\right)\sin \chi_p \,
d\chi_p
\end{equation*}
for the average over the polar angle $\chi_p$ of the 3D Fermi
momentum. Note that our calculations are based on the
first-order approximation in the small parameter $b/\rho$.
According to Eq.~\eqref{Sigma_RI} the symmetrical
${\Sigma_I(-s)=\Sigma_I(s)}$ and antisymmetrical
${\Sigma_R(-s)=-\Sigma_R(s)}$ parts of the off-diagonal self energy
term $\Sigma_2e^{-i\phi}$ can be rewritten as
${\Sigma_R=\Sigma_2e^{-i\phi}s/\rho}$ and
${\Sigma_I=\Sigma_2e^{-i\phi}b/\rho}$.

The self energy obtained above affects the vortex core states in 2D layer in two different ways. The adiabatic part of the induced vortex potential leads to the Andreev localization of QPs with energy smaller than the induced superconducting gap $\Gamma$  within the induced vortex core at distances of the order of $\xi_{2D}$. This forms the CdGM anomalous branch $\epsilon_2(b)$ as in an usual superconductor with the corresponding maximum intrinsic gap $\Gamma$.
Another part of the self energy exponentially decaying at $\rho\sim\xi_S$ contains information about the CdGM states in the bulk SC; it affects the 2D-layer QP behavior at small scales. The adiabatic large-scale part of the self energy
(at $\rho\gg\xi_S$)
is universal; it does not depend on the tunneling models and on possible disorder in the bulk SC, while the short-scale induced vortex potential localized at small distances  does crucially depend on these factors.
Both terms in the induced self energy form the two-scale local DOS (LDOS) radial profile.

\section{Scale separation method}\label{sec-scale-sep-method}

A natural way to solve Eqs.~\eqref{eq-zeta-theta-g-2} is to apply
the scale separation method. We introduce a distance $\rho_0$ satisfying $\xi_S\ll\rho_0\ll\xi_{2D}$
and consider the Green functions in two overlapping spatial
intervals: (i) $\rho\lesssim \rho_0$ and (ii) $\rho \gtrsim \rho_0$.  Next we match the solutions in different spatial domains.

\subsection{Large distances}

At low energies $\epsilon\ll\Delta_\infty$ and large distances $\rho\gg\xi_S$ the induced
vortex potential is given by Eq.~\eqref{fg-asimpt_gen}.
QPs propagating along the trajectories with impact parameters $b>\xi_S$ that miss the
primary vortex core are affected only by this long-distance ($\xi_{2D}\gg\xi_S$)
part of the induced gap potential. In the low energy limit $\epsilon<\Gamma \ll\Delta_\infty$ the appropriate
boundary conditions far from the induced vortex core ($\rho\gg
\xi_{2D}$) are
\begin{equation}\label{bcond-Gamma}
\theta =\frac{\Gamma s/\rho}{\sqrt{\Gamma^2-\epsilon^2}} \ , \;
\zeta =\frac{-\Gamma b/\rho}{\sqrt{\Gamma^2-\epsilon^2}} \ ,
g=\frac{-i\epsilon }{\sqrt{\Gamma ^{2}-\epsilon ^{2}}} \ .
\end{equation}
For both tunneling models and
arbitrary disorder rate inside the superconductor and for $\rho\gg\xi_S$ Eqs.~\eqref{eq-zeta-theta-g-2} take the form:
\begin{subequations}\label{eq-Elen-2D_adiab}
\begin{align}
\hbar v_{2F} \frac{d\zeta}{ds} +2 \epsilon   \theta   -2i g \Gamma s/\rho =0  , \label{new1}\\
\hbar v_{2F} \frac{d\theta }{ds} -2 \epsilon   \zeta  -2i g \Gamma b/\rho=0 , \label{new2}\\
\hbar v_{2F} \frac{d g}{ds} + 2i \theta\Gamma b/\rho +2i\zeta
\Gamma s/\rho =0  \ .\label{new3}
\end{align}
\end{subequations}

The functions $g$ and $\zeta$ are even in $s$ while $\theta$ is
odd, so we can consider only positive $s$ values. We obtain the
solution of the above equations using the first-order perturbation
theory in the impact parameter $b$: $\check w(s)=\check
w_0(s)+\check w_1(s)$, where $ \check w(s) =\left(\zeta, \theta,
ig\right)^T$. As we shall see later, this approximation holds for $|b|\ll \xi_{2D}$. The
zero order in $b$ solution reads
\begin{equation}
\check w_0(s) =\frac{1}{\sqrt{\Gamma^2-\epsilon^2}}\check u_0(s)
+\frac{C}{\sqrt{\Gamma^2-\epsilon^2}}\check u_-(s) \ , \label{g2}
\end{equation}
where
\begin{equation*}
\check u_{\pm}(s)= \left( \begin{array}{c}
\sqrt{\Gamma^2-\epsilon^2}\\ \pm \epsilon \\
 \pm \Gamma\end{array} \right) e^{\pm \lambda s}\ ,\;
 \check u_0(s) = \left(\begin{array}{c} 0\\ \Gamma \\ \epsilon \end{array}\right)
\end{equation*}
and $ \lambda = 2\sqrt{\Gamma^2 -\epsilon^2}/\hbar v_{2F} $.
This solution satisfies the boundary conditions  ${g= -i\epsilon
/\sqrt{\Gamma^2 -\epsilon^2}}$, $\zeta =0$ and ${\theta
=\Gamma/\sqrt{\Gamma^2 -\epsilon^2}}$ for $s \to \infty$ and
$\epsilon^2 <\Gamma^2$.
 The first order correction $\check w_1$ can be written as
\begin{equation}
\check w_1(s) =\frac{C_0(s) \check u_0}{\sqrt{\Gamma^2-\epsilon^2}}  +\frac{C_+(s) \check u_+}{\sqrt{\Gamma^2-\epsilon^2}} +\frac{C_-(s) \check u_-}{\sqrt{\Gamma^2-\epsilon^2}}
  \ , \label{g-corr}
\end{equation}
where
\begin{subequations}\label{C0_C+_C-}
\begin{align}
\xi_{2D}C_0(s)=2 C b \int_{s}^\infty e^{-\lambda s}\frac{ds}{\rho}\ , \label{C0} \\
\xi_{2D}C_+(s)=- b \int_{s}^\infty e^{-\lambda s}\frac{ds}{\rho}\ , \label{C+}\\
\xi_{2D}C_-(s)=- b \int_{s_{c}}^s e^{\lambda s}\frac{ds}{\rho}\ . \label{C-}
\end{align}
\end{subequations}
The lower limit of integration in $C_-$, $s_c$, has to be taken as
$s_c \sim \xi_S$ for trajectories that go through the primary
vortex core, $b\lesssim \xi_S$, so that the logarithmic divergence
is cut off at the distances $\sim \xi_S$ where the long-range
vortex potential $\Sigma_{ad}$ \eqref{Phi} vanishes. For $b \gg \xi_S$ we
have $s_c =0$. The perturbation approach holds as long as $C_0\ll
C$ and $C_+\ll 1$, i.e., as long as $|b|\ll \xi_{2D}$. For
$s\gg\xi_{2D}$ the coefficient $C_0$ decays faster than
exponentially, while
\begin{equation*}
C_+(s)e^{\lambda s} \to C_-(s)e^{-\lambda s} \to -\frac{\Gamma}{2\sqrt{\Gamma^2-\epsilon^2}}\frac{b}{\rho}
\end{equation*}
so that $\zeta$ approaches  $
-(b/\rho)\Gamma/\sqrt{\Gamma^2-\epsilon^2}$ and the corrections to
$\theta$ and $g$ vanish as it should be according to
\eqref{bcond-Gamma}. For a small distance $s=s_0$ defined as $\rho_0^2=s_0^2+b^2$ we have
\begin{subequations}\label{zeta2-theta2-g2-s0}
\begin{align}
\zeta(s_0)&=C+ C_+(s_0)+C_-(s_0), \label{zeta2-s0} \\
\theta(s_0)&=\frac{1}{\sqrt{\Gamma^2-\epsilon^2}}\left\{ \Gamma
-\epsilon C+\Gamma C_0(s_0) \right.\nonumber \\
&+\left. \epsilon [C_+(s_0)-C_-(s_0)] \right\}, \quad \label{theta2-s0}\\
g(s_0)&= \frac{i}{\sqrt{\Gamma^2-\epsilon^2}}\left\{-\epsilon + \Gamma C
-\epsilon C_0(s_0)\right.\nonumber \\
 &- \left. \Gamma [C_+(s_0)-C_-(s_0)]\right\}.\quad \label{g2-s0}
\end{align}
\end{subequations}

\subsection{Matching for large impact parameters}

Far from the primary vortex core at impact parameters $\xi_S\ll b \ll \xi_{2D}$ the perturbation result Eqs. \eqref{eq-Elen-2D_adiab} can be applied along the entire trajectory so that one can put $s_0=s_c =0$. The boundary condition for an odd function requires $\theta (0)=0$.
Since in this case $C_-(0)=0$, we find from  Eq.~\eqref{theta2-s0}
\begin{equation*}
\Gamma +\epsilon C_+(0)=\epsilon C -\Gamma C_0(0) \ .
\end{equation*}
Expressing the coefficients $C_0$ and $C_+$  in terms of the
energy
\begin{equation}
\epsilon=\epsilon_2(b)=\frac{2\Gamma^2 b}{\hbar v_{2F}} \ln
\eta \ ,\label{eps2}
\end{equation}
of bound states in the induced vortex core, with $\eta = \xi_{2D}/|b|$,
$C_0=-2C C_+=C \epsilon_2(b)/\Gamma$, we find
\begin{equation}
C[\epsilon -\epsilon_2(b)]=\Gamma - \epsilon \epsilon_2(b)/2\Gamma
\ .\label{const2}
\end{equation}
  According to Eq.~\eqref{const2}
$\epsilon_2(b)$ is the only spectrum branch in the energy interval
$|\epsilon| \ll \Delta_\infty$. The Green function is
\begin{multline}
g(s)=\frac{-i\epsilon}{\sqrt{\Gamma^2-\epsilon^2}}
 + \frac{i\Gamma C}{\sqrt{\Gamma^2-\epsilon^2}}e^{-\lambda s}
-\frac{i\epsilon C_0(s)}{\sqrt{\Gamma^2-\epsilon^2}} \\
-\frac{i\Gamma }
{\sqrt{\Gamma^2-\epsilon^2}}\left[C_+(s)e^{\lambda s} -C_-(s)e^{-\lambda s}\right]
\label{g2-s-highb} \ .
\end{multline}
For $s\gg \xi_{2D}$ we have $C_0 \to 0$, $C_+ e^{\lambda s}-C_-
e^{- \lambda s}\to 0$, so that the first term is the homogeneous
background while the rest terms describe the vortex contribution.
To obtain the retarded function  for $\epsilon^2 >\Gamma^2$ one
has to continue $\sqrt{\Gamma^2-\epsilon^2} $ analytically
throughout the upper half-plane of complex $\epsilon$ keeping $\Re
\sqrt{\Gamma^2 -\epsilon^2}>0$.

\subsection{Matching for small impact parameters}

To find the Green functions for
small impact parameters $b\lesssim\xi_S$ one has to match Eqs.~\eqref{zeta2-theta2-g2-s0} with the solution obtained in the vortex core region.
For small $s<s_0$ we assume that the even parts of the Green function
$g(s)$ and $\zeta(s)$ are nearly constant in the interval
$0<s<s_0$. Integrating
Eq.~\eqref{eq-theta-2} over $s$ from $0$ to
$s_0$ along the trajectory we find the matching condition
\begin{equation}
\frac{\hbar v_{2F}}{2}\theta (s_0)=\zeta(s_0)\int_0^{s_0} \Sigma_1\,
 ds + i g(s_0)\int_0^{s_0} \Sigma_I\, ds  \ .\label{bcond-s0}
\end{equation}
Equation~\eqref{bcond-s0} determines the
constant $C$. Its poles  define the eigenstates
of excitations as
functions of energy and the impact parameter.
While deriving the effective boundary condition \eqref{bcond-s0}
for $b\lesssim\xi_S$, one needs to separate the exponentially
converging parts $\Sigma_{1,I}^{loc}$ at $s\sim\xi_S$ from the
long-distance, $s\gg \xi_S$, asymptotics of $\Sigma_{1,I}$.
For ${\epsilon \ll
\Delta_\infty}$ the
long-distance expressions, Eq.~\eqref{fg-asimpt_gen}, 
yield $\Sigma_1\to 0$,
$\Sigma_I \to \Gamma b/\rho$. Therefore,
\begin{multline}\label{Sigma_I_int}
\int_0^{s_0} \Sigma_I\, ds =\int_0^{\xi_S} \Sigma_I^{loc}\,
 ds +\Gamma\int_{\xi_S}^{s_0} b/\rho \, ds \\
 \approx \int_0^{\infty} \Sigma_I^{loc}\, ds +\Gamma b \ln({s_0}/{\xi_S}) \ .
\end{multline}
while $\int_0^{s_0} \Sigma_1\, ds$ can be extended to infinity.
The localized self energy  parts $\Sigma_{1}$,  $\Sigma_{I}^{loc}$ determine
the small-distance LDOS and spectrum of excitations
and depend on the particular tunneling mechanism.

\section{ Multiple vortex core in the clean limit.
 Quasiparticle spectrum and density of states.}\label{sec-ind-vort}

\subsection{Isotropic Fermi surface} \label{sec-ind-vort:subsec-clean}

In this section we consider an idealized picture
without any disorder. For large impact parameters, $b\gg \xi_S$, the corresponding
solutions for the Green functions, Eq. (\ref{g2-s-highb}), coincide with the standard CdGM
expressions where the gap value is replaced with $\Gamma$. The corresponding
anomalous spectrum for 2D excitations is given by Eq.~\eqref{eps2}.\cite{KramerPesch,Kopnin-book}
This modified CdGM branch dominates in the
LDOS at large distances $\rho\gg\xi_S$.

The normalized LDOS is defined as an average over the
trajectories:
\begin{equation*}
N({\bf
r},\epsilon)=\int_0^{2\pi}N_\epsilon(s,b)\frac{d\alpha^\prime}{2\pi} =
\int_{-\rho}^{\rho}\frac{N_\epsilon(\sqrt{\rho^2-b^2},
b)} {\sqrt{\rho^2-b^2}}\frac{db}{\pi}
\end{equation*}
where
$
N_\epsilon (s,b)= \left[g^R(s,b)-g^A(s,b)\right]/2
$, $s=\rho \cos\alpha^\prime$, and $b=-\rho\sin\alpha^\prime$.
For $|\epsilon| <\Gamma$, a nonzero LDOS comes only from the
vortex contribution of the second and third terms in
\eqref{g2-s-highb} due to the presence of a pole in the
coefficient $C$ according to Eq.~\eqref{const2}.
The Green
functions and LDOS reach their long-distance values
$g=-i\epsilon/\sqrt{\Gamma^2-\epsilon^2}$ and
$N=\Re|\epsilon|/\sqrt{\epsilon^2-\Gamma^2}$
as $\rho\to \infty$. For $\rho\gg\xi_S$ the trajectories with
large impact parameters $b\gtrsim\xi_S$ give the main contribution
to the LDOS. In the region ${\xi_S\ll\rho\ll\xi_{2D}}$ we get the
angle--resolved density of states in the form:
\begin{eqnarray}
N_\epsilon(s, b)&=& \frac{\sqrt{\Gamma^2 -\epsilon^2}(\Gamma^2 -\epsilon^2/2)}{\Gamma^2}\nonumber \label{dos<}\\
&&\times \pi \delta[ \epsilon -\epsilon_2(b)]\ , \phantom{223i space} |\epsilon
|<\Gamma \\
N_\epsilon(s, b)&=&\frac{\sqrt{\epsilon^2-\Gamma^2}[\Gamma^2
-\epsilon_2^2(b)/2]}{{\rm sign}(\epsilon)\Gamma^2[\epsilon -\epsilon_2(b)]} \ , \;
|\epsilon |>\Gamma\ . \label{dos>}
\end{eqnarray}
Thus, the corresponding LDOS in the energy interval $|\epsilon|
<\Gamma$ has the only peaks at $\epsilon=\epsilon_2(\pm\rho)$:
\begin{multline}
N(\rho, \epsilon)=
\frac{1}{\pi}\int\limits_{-\rho}^{\rho}N_\epsilon(\sqrt{\rho^2-b^2},
b) \frac{db}{\sqrt{\rho^2-b^2}}=\\= \Re\frac{\sqrt{\Gamma^2
-\epsilon^2}(1 -\epsilon^2/2\Gamma^2)}
{\sqrt{\epsilon_2^2(\rho)-\epsilon^2}} \ .
\end{multline}
For energies above the induced gap, $|\epsilon|
>\Gamma $, for the same distances the LDOS is monotonically increasing with
$|\epsilon|$
to its normal state value:
\begin{equation}
N(\rho, \epsilon)= \sqrt{\epsilon^2-\Gamma^2}\left[\frac{|\epsilon|}
{2\Gamma^2}+\frac{(1 -\epsilon^2/2\Gamma^2)}{\sqrt{\epsilon^2-\epsilon_2^2(\rho)}}\right]\ .
\end{equation}

A trajectory with a small impact parameter $b\lesssim \xi_S$ can be divided into the part far from the primary vortex core, and the region inside the core. Far from the core the solution is found using the vortex potentials Eq.~\eqref{fg-asimpt_gen}.
The self energies of the
primary vortex in Eq.~\eqref{eq-Eliashb-st} have poles at the usual  CdGM energy $\epsilon_0(b)$ with the corresponding wave functions exponentially localized within $\rho\sim\xi_S$ and regular parts extending over large distances $\rho\to\pm\infty$ \cite{KramerPesch,Kopnin-book}:
\begin{equation}
\Sigma_R=\Gamma \theta_S\ , \;
\Sigma_I=-\Gamma \zeta_S \ .
\end{equation}

Note that the localized part $\Sigma_2^{loc}$ of the effective
order parameter $\Sigma_2$ has the coordinate dependence
$\Sigma_2^{loc}=i\Sigma_I^{loc}(b,s)e^{i\alpha}$ with \emph{zero}
circulation, unlike its adiabatic part \eqref{fg-asimpt_gen}
$\Sigma_2(\rho\gg\xi_S)=\Gamma e^{i\phi}$. As we will see below it
is this different angular dependence of the effective gap
asymptotics, which leads to the formation of a ``shadow'' of the
bulk SC anomalous branch in the excitation spectrum and LDOS in
the 2D layer.

\begin{figure}[t]
\includegraphics[width=0.80\linewidth]{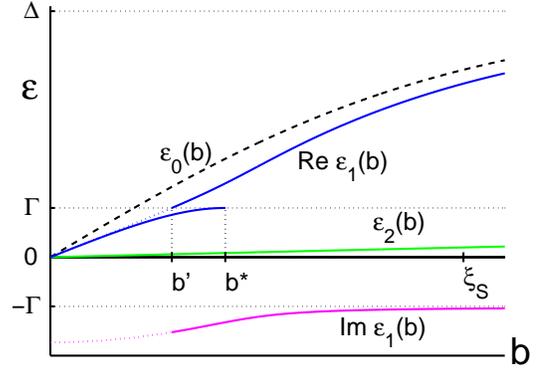}
\caption{(Color online) Two localized branches, $\epsilon_1(b)$ and $\epsilon_2(b)$ of the spectrum, Eq.~\protect\eqref{pole_coh}, in the limit of coherent tunneling, for $\epsilon<\Gamma$.
$b^*$ is defined as $\epsilon_1(b^*)=\Gamma-0$, while
$b'$ corresponds to $\Re\epsilon_1(b')=\Gamma+0$.
The spectrum satisfies $\epsilon_{1,2}(-b)=-\epsilon^*_{1,2}(b)$.} \label{fig-spectrum2}
\end{figure}
Using Eqs.~\eqref{C0_C+_C-} for the long-distance part of the trajectory we find
\begin{subequations} \label{C0C+C-}
\begin{align}
C_0(s_0) &= \frac{2 C b }{\xi_{2D}} \ln \frac{1}{\lambda s_0} \ , \\
C_+(s_0)\pm C_-(s_0)&\approx -\frac{b}{\xi_{2D}} \ln
\frac{1}{\lambda\xi_S}\approx-\frac{\epsilon_2(b)}{2\Gamma} \ .
\end{align}
\end{subequations}
We now match the asymptotic solution
Eqs.~\eqref{zeta2-theta2-g2-s0} obtained for $s\geq s_0$ with the solution for the short-distance part of the trajectory, Eqs. (\ref{sigma-coherent}) and (\ref{zeta}) - (\ref{K}), using Eq.~\eqref{bcond-s0} and
Eq.~\eqref{Sigma_I_int}.
As a result,
\begin{multline}
C\left[ \xi_{2D}[\epsilon-\epsilon_2(b)] +
2[\Gamma- \sqrt{\Gamma^2-\epsilon^2}-
\frac{\epsilon \epsilon_2(b)}{\Gamma}]\int_0^\infty \zeta_0\, ds \right] \\
= \left[\xi_{2D}\Gamma +2\epsilon \int_0^\infty \zeta_0\,
 ds - \xi_{2D} \frac{\epsilon \epsilon_2(b)}{2\Gamma}\right.
 \\
\left.  -(\Gamma +\sqrt{\Gamma^2-\epsilon^2})\frac{\epsilon_2(b)}
{\Gamma}\int_0^\infty \zeta_0 \, ds \right]  \ ,\label{const}
\end{multline}
where $\zeta_0(s)$ is the localized part of $\zeta_S$ and
\begin{equation}
\int_0^\infty \zeta_0\, ds =\frac{\hbar v_\parallel}{2[\epsilon -\epsilon_0(b)]}\ .\label{int-zeta}
\end{equation}
Here we put $g=i\zeta_0$ and replace the cutoff parameter in
\eqref{eps2} by $\eta=\xi_{2D}/\xi_S$. For $b \gg \xi_S$ the contributions
from  the primary vortex core
proportional to $\int_0^\infty \zeta_0\, ds$ vanish since the
trajectory misses the core, and Eq.~\eqref{const} goes over into Eq.~\eqref{const2}.

For small $b\ll\xi_{2D}$ the Green function has a pole when
\begin{multline}
P(\epsilon,b)=[\epsilon-\epsilon_2(b)][\epsilon -\epsilon_0(b)]
\\+ q_v \left[\Gamma^2 -
\Gamma \sqrt{\Gamma^2-\epsilon^2}-\epsilon \epsilon_2(b)\right] =0 \label{pole1}
\end{multline}
where $q_v= v_\parallel/v_{2F}$.
 One can show that including the higher order terms in the parameter
 $\epsilon_2(b)/\Gamma$
 the corresponding energy dispersion relation takes the form:
 \begin{equation}
\frac{[\epsilon-\epsilon_2(b)][\epsilon -\epsilon_0(b)]}{\Gamma q_v} + \Gamma - \sqrt{\Gamma^2-[\epsilon-\epsilon_2(b)]^2} =0 \label{pole_coh}
\end{equation}
\begin{figure}[t]
\includegraphics[width=0.8\linewidth]{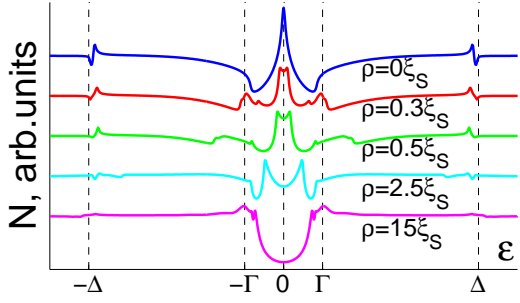}
\caption{(Color online) LDOS in logarithmic scale for coherent
 tunneling in the
clean limit. Curves, taken for different distances $\rho$ from the vortex center,
are vertically shifted for clarity. The peaks in LDOS exist up to distances $\sim \xi_{2D}$.
Here $\Delta/\Gamma=5$, $q_v=1$.}
\label{fig-DOS-Coher}
\end{figure}
 For $b \lesssim \xi_S$, the cutoff parameter in Eq.~\eqref{eps2}  should be replaced with
$\eta=\xi_{2D}/\xi_S$.

The resulting two-scale spectrum is shown in
Fig.~\ref{fig-spectrum2}.   There are two real-valued
branches in the range $|\epsilon|<\Gamma$ crossing zero of energy as functions of the
impact parameter and one complex-valued branch in the range $\Gamma<|\epsilon|<\Delta_\infty$. The
lowest-energy branch  $\epsilon_2(b)$  has a scale $\xi_{2D}$ as a function of the impact parameter: For
$b\lesssim \xi_{2D}$ it is given by Eq.~\eqref{eps2} with the proper cutoff parameter $\eta$ as discussed above and saturates at $\epsilon=\Gamma$ for
$b>>\xi_{2D}$. The branch $\epsilon_1(b)$ has a scale $\xi_S$:
For $\epsilon<\Gamma$ it goes slightly below the CdGM spectrum $\epsilon_0(b)$
of the bulk SC,
 $\epsilon_1(b) =(1+q_v/2)^{-1}\epsilon_0(b)$.
Above $\Gamma$ the spectrum
transforms into a scattering resonance due to the decay into
delocalized modes propagating in the 2D layer: $\epsilon_1(b)
=\epsilon_0(b)-i\Gamma q_v$ for $|\epsilon|\gg\Gamma$.
Since Eq.~\eqref{pole_coh} determines a pole of the retarded
Green function in the lower half-plane of complex $\epsilon$,
the square root in Eq.~\eqref{pole_coh} should be analytically
continued through the cut going from $-\infty$ to $-\Gamma$ and
from $\Gamma$ to $+\infty$. As a result, $\epsilon_1(b)$ has a
discontinuity at $\epsilon_1=\Gamma$ with $b^\prime/\xi_S\approx 0.29  $ and
$b^*/\xi_S \approx 0.42$.

The two branches appear due to the presence of two sub-systems, the
bulk SC and the 2D proximity layer, each with its own anomalous
branch. The existence  of two anomalous branches follows also from the
index theorem \cite{volovik,Shiozaki12}. Indeed, its application requires that
both zero of the quasiclassical Hamiltonian at the Fermi surface
and its singularity at $\epsilon =\epsilon_0(b)$ are taken into
account when calculating the topological invariant. As a result,
the number of anomalous branches is increased up to 2 for a
single-quantum vortex.

The multiple-branch spectrum results in multiple-peak structure in the
LDOS (Fig.~\ref{fig-DOS-Coher}), which
appears to be most pronounced deeply
inside the primary core (at distances $\rho \lesssim
\xi_S^2/\xi_{2D}$ when $\epsilon_1<\Gamma$) thus illustrating
the two-scale structure of the vortex core.
The LDOS is obtained from the angle-resolved DOS (normalized by its normal state value)
${N_\epsilon(s,b)= [g^R(s,b)-g^A(s,b)]/2}$ averaged over the trajectory direction.
%

The angle-resolved DOS for small energies $|\epsilon|\ll\Gamma$
and $\rho\lesssim\xi_S$ reads
\begin{multline}
N_\epsilon(s,b)=\frac{\pi\Gamma q_v}{2}\delta[\epsilon-\epsilon_1(b)]+\frac{\pi\Gamma(q_v+2)}{2}\delta[\epsilon-\epsilon_2(b)] \ .
\end{multline}
Here we neglect the terms $\epsilon\epsilon_2(b)/\Gamma^2$ and
$\epsilon_2(b)/\epsilon_1(b)$ and put
$\epsilon_0(b)/\epsilon_1(b)=1+q_v/2$ according to low energy asymptotics.
 In this case  the LDOS
\begin{multline}\label{LDOS_Coh_small_E}
N(\rho,\epsilon)=\Re\frac{\Gamma q_v}{2\sqrt{\epsilon_1^2(\rho)-\epsilon^2}}+\Re
\frac{\Gamma (q_v+2)}{2\sqrt{\epsilon_2^2(\rho)-\epsilon^2}}
\end{multline}
reveals a two-peak structure vs energy at
$\epsilon=\epsilon_{1,2}(\rho)$. For $|\epsilon| \sim \Gamma$, one
can neglect $\epsilon_2(b)$ and obtain:
\begin{equation}
[\epsilon -\epsilon_0(b)]\left[\Gamma +
\sqrt{\Gamma^2-\epsilon^2}\right] + q_v \Gamma \epsilon =0 \ . \label{pole2}
\end{equation}
For $|\epsilon| >\Gamma$ the dispersion relation is complex valued
and for retarded functions takes the form:
\begin{equation}
\epsilon [\epsilon -\epsilon_0(b)] +
 q_v\Gamma \left[\Gamma   +i {\rm sign}(\epsilon)
 \sqrt{\epsilon^2- \Gamma^2}\right]  =0  \ .\label{E>G}
\end{equation}
The latter equation describes the resonant states in the 2D vortex
core which decay into the QP waves propagating in the
2D layer above the induced gap.

Finally, the whole spectrum  structure, shown in
Fig.~\ref{fig-spectrum2}, has  two anomalous branches: (i) one of
them $\epsilon_2(b)$ is completely real-valued and follows the
CdGM spectrum for the superconductor with homogeneous gap
$\Gamma$; (ii) another one is close to the bulk CdGM spectrum, but
has a discontinuity at $\epsilon=\Gamma$, where it becomes
essentially complex.

Thus, the LDOS for energies above the induced gap $|\epsilon|>\Gamma$ and small distances $\rho, b\lesssim\xi_S$ reads
%
\begin{multline}\label{LDOS_Coh_big_E}
N(\rho ,\epsilon) =\frac{\sqrt{\epsilon^2-\Gamma^2}}{|\epsilon|} + \frac{q_v\Gamma^2}{2|\epsilon|}\times\\
\Re\frac{\sqrt{\epsilon^2-\Gamma^2}-i\Gamma}{\sqrt{(\epsilon^2+q_v\Gamma^2+i q_v\Gamma\sqrt{\epsilon^2-\Gamma^2})^2-\epsilon^2\epsilon_0^2(\rho)}}
\end{multline}
and has the only peak at $\epsilon=\Re\epsilon_1(\rho)$ of the
height $\sim\Gamma^2/\epsilon_0^2(\rho)$ for
$\rho\gtrsim\xi_S^2/\xi_{2D}$. In the opposite limit of rather
large distances $\rho>\xi_S^2/\xi_{2D}$ at $|\epsilon| > \Gamma$,
the spectrum reduces to the CdGM spectrum with a finite
broadening:
\begin{equation}
\epsilon_1(b) =\epsilon_0(b) -i\Gamma q_v\label{E>>G} \ .
\end{equation}
The LDOS has a small difference from its normal state value
$N_0=1$:
\begin{equation}
N(\rho ,\epsilon) =1+ \frac{q_v\Gamma^2}{2\epsilon^2}\Re\frac{|\epsilon|-i\Gamma}{\sqrt{(\epsilon+i q_v\Gamma)^2-\epsilon_0^2(\rho)}}
\end{equation}

The LDOS in the whole energy range~(\ref{LDOS_Coh_small_E},
\ref{LDOS_Coh_big_E}) has two or even three peaks for such
distances. The latter case is realized at the distances
corresponding to $b'<b<b^*$, where the spectrum vs the impact
parameter has 3 anomalous branches.

The numerical LDOS patterns have been
obtained by the subsequent solving of the two sets of Eilenberger
equations in the Riccati parametrization \cite{Schopohl}: first, we
calculate the Green functions in the bulk
SC using the approximation $\Delta_0(\rho)=\Delta_\infty\rho/\sqrt{\rho^2+\xi_S^2}$
and next we solve Eq.~\eqref{eq-Eliashb-st} in the 2D layer using Eq.~\eqref{selfen-coh}.

\subsection{Anisotropic Fermi surface} \label{sec-anisotropic}

\begin{figure}[t]
\includegraphics[width=0.9\linewidth]{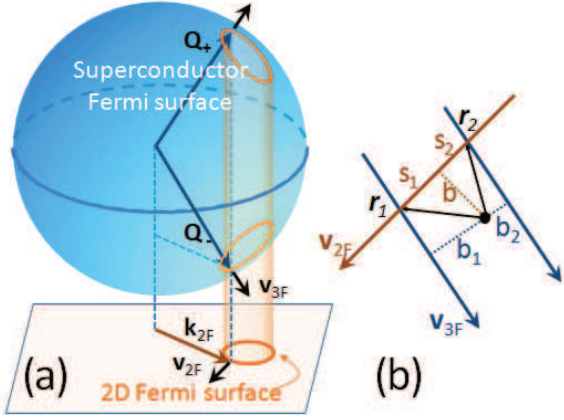}
\caption{(a) An example of anisotropic Fermi surfaces showing a spherical 3D Fermi surface on top of a part of a 2D Fermi surface in the layer shifted from the center of its Brillouin zone. The closed loops show the 2D Fermi line and its projections onto the 3D Fermi surface. The directions of the 3D Fermi velocity projection ${\bf v}_{3F}$ on the plane $z=0$ in the bulk does not coincide with that in the 2D layer, ${\bf v}_{2F}$. (b) Different points ${\bf r}_1$ and ${\bf r}_2$ specified by $s_1 $ and $s_2$ on a 2D trajectory with given impact parameter $b$ belong to trajectories in 3D with different impact parameters $b_1$ and $b_2$.} \label{fig-anisotr}
\end{figure}

Here we briefly discuss the effects of anisotropic Fermi surfaces in 3D and/or 2D systems. We will be interested only in main distinctions which the anisotropy causes within the coherent tunneling model as compared to the isotropic case considered above. For anisotropic surfaces, one can also apply the method of scale separation in the same manner as we did it in Sec. \ref{sec-scale-sep-method}. The consideration for the region of large impact parameters does not differ significantly, such that the solution for the Green functions together with the matching conditions look similar to Eqs. (\ref{g2}), (\ref{C0_C+_C-}), (\ref{C0C+C-}) and (\ref{bcond-s0}), (\ref{Sigma_I_int}). However, the region of small impact parameters of the order of $\xi_S$ gives an essentially different result. The main distinction is that the directions of QP trajectories which are determined by the group velocities $\partial \epsilon_{2D}/\partial {\bf k}$ and $\partial \epsilon_{3D}/\partial {\bf Q}$ for given in-plane momentum in 2D and 3D systems do not coincide, Fig.~\ref{fig-anisotr}(a). As a result, the integral in Eq. (\ref{bcond-s0}) along a 2D trajectory involves trajectories with different impact parameters used to parameterize the 3D Green functions, see Fig.~\ref{fig-anisotr}(b). Within the quasiclassical approximation, the integral will thus give an imaginary part which comes from the delta function at the 3D core spectrum and a real contribution from a smooth dependence. The spectrum $\epsilon_2(b)$ at small impact parameters thus becomes broadened and shifted from its initial position. The imaginary contribution appears 
due to the coupling of the QP trajectory in the 2D layer with a quasiclassical continuum of trajectories inside the superconductor corresponding to different impact parameters. This coupling results from the non-conservation of the angular momentum in the anisotropic system. The imaginary contribution will be also present if the primary-core spectrum is broadened by disorder or inelastic scattering. The situation is in many respects similar to that for incoherent tunneling model discussed in the following section. Of course, attributing the origin of the imaginary part of energy for the anisotropic case to the continuum of states in the bulk SC, we ignore the angular momentum quantization in the primary vortex core. The true quantum mechanical consideration accounting for the level quantization could possibly change this conclusion and lead to a real-valued energy spectrum for ideal systems without disorder.  


In this Section we consider the low-energy behavior of the Green function at small impact parameters, $b\ll \xi_{S}$ where the spectral energy $\epsilon_2(b)$ is very small and can be neglected.
We assume that the trajectories in the bulk SC and in the 2D layer do not coincide; the Fermi velocities ${\bf v}_{2F}$ and ${\bf v}_{3F}$ are at an angle $\delta \alpha$ to each other, see Fig.~\ref{fig-anisotr}(b). The impact parameter $b_S$  and trajectory coordinate $s_S$ in the superconductor are coupled with the ones in the 2D layer ($b$ and $s$)
through
\begin{eqnarray*}
b_S=\rho \sin(\phi -\alpha +\delta\alpha) = b\cos\delta\alpha +s\sin\delta\alpha \ ,\\
s_S=\rho \cos(\phi -\alpha +\delta\alpha) = s\cos\delta\alpha -b\sin\delta\alpha \ .
\end{eqnarray*}
As we know,
at small distances $\rho\ll \xi_S$ the part $\zeta_S$ of anomalous Green function $f_S$ in the bulk superconductor is large compared with $\zeta_S\gg\theta_S$. Neglecting the latter, we have for the self energies
in Eq. \eqref{Sigma_RI}:
\begin{subequations}\label{Sigma_RI_dalph}
\begin{align}
\Sigma_R=\Sigma_1 \sin\delta\alpha ,\quad\\
\Sigma_I=\Sigma_1 \cos\delta\alpha ,\quad
\end{align}
\end{subequations}
The diagonal self energy is $\Sigma_1=i\Gamma g_S\approx -\Gamma\zeta_S$. Note that the self energies depend on trajectory coordinate $s$ in 2D layer through the impact parameter $b_S=b\cos \delta\alpha + s\sin\delta\alpha$ in the bulk superconductor and do not possess definite symmetry in $s$. Therefore, one needs to consider the region inside the primary core more carefully allowing for contributions from even and odd components of the corresponding functions.

As in Sec. \ref{sec-scale-sep-method} we use the scale separation method and subdivide a 2D layer trajectory with a small impact parameter $b\lesssim \xi_S$ into the long-distance part far from the primary vortex core, and the region inside the core. We introduce a distance $\rho^\prime$ satisfying $\xi_S\ll\rho^\prime\ll\xi_{2D}$
and consider the Green functions in two overlapping spatial
intervals: (i) $\rho\lesssim \rho^\prime$ and (ii) $\rho \gtrsim \rho^\prime$.  Next we match the solutions in different spatial domains. Far from the core the solution is found using the vortex potentials Eq.~\eqref{fg-asimpt_gen}.

In the region inside the primary vortex core the self energies play the most important role.
Using the approximation \eqref{Sigma_RI_dalph}, $\Sigma_R=\Sigma_1 \sin\delta\alpha$, $\Sigma_I=\Sigma_1 \cos\delta\alpha$
and neglecting $\epsilon$,
Eqs.~\eqref{eq-zeta-theta-g-2} at small distances $s<s_0$ can be written in the matrix form
\begin{equation}\label{eq-zeta-theta-g-small}
\frac{d}{ds}\check w+\frac{2\Sigma_1}{\hbar v_{2F}}\check A \check w = 0 \ .
\end{equation}
As in Sec.~\ref{sec-scale-sep-method} we use the vector $\check w =(\zeta, \, \theta , \, ig)^T$. The constant matrix
$$\check A = \begin{pmatrix}0 & 1 & -\sin\delta\alpha\\ -1 & 0 & -\cos\delta\alpha \\ -\sin\delta\alpha & -\cos\delta\alpha & 0\end{pmatrix}$$
has threefold degenerated zero eigenvalue, therefore the solution of Eq.~\eqref{eq-zeta-theta-g-small} can be written in terms of mutually orthogonal eigenvector $\check v_0=\left(-\cos\delta\alpha, \sin\delta\alpha, 1\right)^T$ and adjoined vectors $\check v_1=\left(-\sin\delta\alpha, -\cos\delta\alpha, 0\right)^T$ and $\check v_2=\left(\cos\delta\alpha, -\sin\delta\alpha, 1\right)^T$:
\begin{equation} \label{w-expansion}
\check w=C_0(s)\check v_0+C_1(s)\check v_1+C_2\check v_2 \ ,
\end{equation}
where $\check A \check v_0=0$, $\check A\check v_1=\check v_{0}$, and $\check A\check v_2=2\check v_{1}$. Therefore
\begin{subequations}
\begin{align}
\frac{d C_1(s)}{dx} &= -2\frac{2\Sigma_1(s)}{\hbar v_{2F}}C_2 \ ,\\
\frac{d C_0(s)}{dx} &= -\frac{2\Sigma_1(s)}{\hbar v_{2F}}C_1(s) \ .
\end{align}
\end{subequations}
The solution is
\begin{eqnarray}
C_1=C_1^0+2C_2 I(s)\ , \; C_0=C_0^0+C_1^0 I(s)+ C_2 I(s)^2, \; \label{solution_small_s}
\end{eqnarray}
where $\Sigma_1=-\Gamma\zeta_S$ and
\begin{equation}\label{int_zetaS}
I(s)=\frac{2\Gamma}{\hbar v_{2F}}\int_0^s\zeta_S(s') ds' \ .
\end{equation}
The three equations (\ref{eq-zeta-theta-g-small}) are not independent because of the normalization $g^2+\zeta^2+ \theta^2=1$. Therefore, the three coefficients $C_1^0$, $C_2$, and $C_0^0$  are coupled through
$
(C_1^{0})^2 =4C_2C_0^0
$.

Multiplying Eq. \eqref{w-expansion} by the vectors $\check v_2^+$ and $\check v_1^+$ and using Eq. (\ref{solution_small_s}) one can obtain 4 conditions at $s=\pm s_0$
\begin{subequations}\label{bound_cond_gen}
\begin{align}
\left(\zeta\cos\delta\alpha-\theta\sin\delta\alpha+i g\right)_{\pm s_0} = C_2 \ ,\\
\left(\zeta\sin\delta\alpha+\theta\cos\delta\alpha\right)_{\pm s_0} = -C_1^0-2C_2 I(\pm s_0)
\end{align}
\end{subequations}

Excluding the coefficients $C_2$ and $C_1^0$ we find
\begin{subequations}\label{bound_cond_odd}
\begin{eqnarray}
&&\left[\zeta\cos\delta\alpha-\theta\sin\delta\alpha+i g\right]_{s_0} =0 \ , \label{bound_cond_zeta}\\
\label{bound_cond_theta}
&&\left[\zeta\sin\delta\alpha +\theta\cos\delta\alpha\right]_{s_0}  \nonumber \\
&& \qquad \qquad + 2I_{odd}\left\{\zeta\cos\delta\alpha-\theta\sin\delta\alpha+i g\right\}_{s_0}=0  ,\; \qquad
\end{eqnarray}
\end{subequations}
where the integral $I(s)=I_{even}(s)+I_{odd}(s)$ in Eq.~\eqref{int_zetaS} separated into even $I_{even}(-s)=I_{even}(s)$ and odd $I_{odd}(-s)=-I_{odd}(s)$ parts and $\left[x\right]_{s_0} = x(s_0)-x(-s_0)$ and $\left\{x\right\}_{s_0} = x(s_0)+x(-s_0)$. The integral
\[
I_{odd}=\frac{\Gamma}{\hbar v_{2F}}\int_{-s_0}^{s_0}\zeta_S(s^\prime)\, ds^\prime
\]
takes the form
\begin{eqnarray}
I_{odd}=
\frac{\Gamma v_{\parallel }}{v_{2F}2\Lambda \sin\delta\alpha} \mathcal{P}\int_{-\infty}^\infty
\frac{e^{-K(z\cot\delta\alpha -b\sin\delta\alpha)}}{\left[ \epsilon -\epsilon
_{0}(b\cos\delta \alpha +z)\right]} \, dz \qquad \nonumber \\
\mp\frac{i \pi  v_{\parallel }\Gamma}{2 \Lambda v_{2F}\Omega \sin\delta \alpha} \exp\left[-K\left(\frac{\epsilon \cot \delta\alpha}{\Omega }-\frac{b}{\sin \delta\alpha}\right)\right] \ , \qquad \label{Iodd}
\end{eqnarray}
where we put $s\sin\delta\alpha =z$. The second term comes from the delta-function contribution at one of the primary core states, see Eq. (\ref{zeta}); the upper (lower) sign corresponds to retarded (advanced) function. For $\delta\alpha \lesssim \epsilon/\Delta$ the second term disappears while the first gives the real pole contribution which is equivalent to Eq. (\ref{int-zeta}). One concludes that the imaginary part disappears only for trajectories which are almost parallel (within an angle $\delta\alpha \lesssim \epsilon/\Delta$). For $\delta \alpha \gg \epsilon/\Delta $ the first (real) term vanishes since the expression under the integral becomes odd in $z$. Note that for $\epsilon =0$ and $b=0$ the real term vanishes exactly.

Equations (\ref{bound_cond_odd}) are the matching conditions with the solution in the large-distance region, $s>s_0$. They are generalizations of the matching condition Eq. \eqref{bcond-s0} derived earlier for the isotropic situation. The two conditions Eqs. (\ref{bound_cond_odd}) determine the even and odd parts of the Green functions.

The long-distance solution is found in the same way as in Sec. \ref{sec-scale-sep-method}. However, it does no longer have a definite symmetry with respect to $s \to -s$. We separate the even and odd components $\check w =
\check w_{even}+ \check w_{odd}$ and consider both $s>0$ and $s<0$. In this Section we only discuss the behavior of the Green function for low energies and small impact parameter. We thus neglect the corrections to $\check w$ proportional to $b/\xi_{2D}$. In this case $\check w_{even}$ is given by Eq. (\ref{g2}) where now
\begin{equation*}
\check u_{\pm}(s)= \left( \begin{array}{c}
\sqrt{\Gamma^2-\epsilon^2}\\ \pm \epsilon {\rm sign}(s) \\
 \pm \Gamma\end{array} \right) e^{\pm \lambda |s|}\ ,\;
 \check u_0(s) = \left(\begin{array}{c} 0\\ \Gamma {\rm sign}(s)\\ \epsilon \end{array}\right)
\end{equation*}
and
\begin{equation}\label{w-odd}
\check w_{odd}=\frac{\tilde C {\rm sign}(s)}{\sqrt{\Gamma^2-\epsilon^2}}\check u_- (s)
\end{equation}

Equation \eqref{bound_cond_zeta} gives
\begin{equation}
\tilde C = \frac{\sin\delta\alpha(\Gamma - C\epsilon)}{\sqrt{\Gamma^2-\epsilon^2}\cos\delta\alpha - \Gamma} \label{tildeC}
\end{equation}
Using Eqs. (\ref{g2}), (\ref{w-odd}), and (\ref{tildeC}) we find the combinations
$\zeta(s_0)\pm \zeta(-s_0)$, $\theta(s_0)\pm \theta(-s_0)$, and $ig(s_0)+ig(-s_0)$ in terms of the coefficient $C$. Next we insert these combinations into Eq. (\ref{bound_cond_theta}) and find
\begin{eqnarray}
C[\epsilon- 2ZI_{odd}]= \Gamma +2\epsilon I_{odd}\cos\delta\alpha \label{C-anisotr}
\end{eqnarray}
where
\begin{equation}
Z=\frac{[\sqrt{\Gamma^2-\epsilon^2} \cos \delta\alpha -\Gamma]^2-\epsilon^2 \sin^2\delta\alpha}{ \sqrt{\Gamma^2-\epsilon^2} -\Gamma \cos\delta\alpha }\label{Z-anisotr}
\end{equation}
Equations (\ref{C-anisotr}), (\ref{Z-anisotr}) are the counterparts of Eq. (\ref{const2}) for an asymmetric case and transform into it for $\delta\alpha \to 0$.

For $\epsilon \ll \Gamma$ we have
$
Z=1-\cos\delta\alpha
$.
For $\delta\alpha \gtrsim \epsilon/\Delta $ the integral $I_{odd}$ Eq. (\ref{Iodd}) has only imaginary part. Therefore,
\begin{eqnarray}
C=\frac{\Gamma }{\epsilon -\epsilon_2(b)\pm i \gamma} \label{C-zeroE}
\end{eqnarray}
where
\begin{eqnarray}
\gamma = \frac{\pi v_\parallel \Gamma \tan(\delta\alpha/2)}{\Lambda v_{2F}\Omega } e^{-K(\rho_0)}\sim \Gamma^2/\Delta \ ,
\end{eqnarray}
and $\rho_0=|b/\sin \delta\alpha |$. In Eq. (\ref{C-zeroE}) we include the energy $\epsilon_2(b)$ which can be obtained by more detailed calculations taking into account corrections due to $b/\rho$ in the same way as in Section~\ref{sec-scale-sep-method}. The function $K(\rho_0)$ decays exponentially as $e^{-\rho_0/\xi_S}$ for impact parameters larger than the primary core size, $b \gtrsim \xi_S$.

Therefore, the imaginary term in \eqref{C-zeroE} does not disappear unless $\delta\alpha$ is very small. It results in smearing of the adiabatic energy level $\epsilon_2(b)\ll \Gamma$ and in a Lorentzian behavior of the DOS due to tunneling into the primary vortex core states. We remind that this result is obtained within the quasiclassical approximation.


\section{Disorder effects.}\label{sec-ind-vort:subsec-disorder}

\subsection{Multiple core.\ Clean limit with incoherent tunneling.}

We study the disorder effects by
introducing the momentum scattering first into the tunneling
process as described by the incoherent tunneling model.
Since the tunneling is considered as a perturbation  one can assume a
specular QP scattering at the interface on the bulk side and, thus, use
the results of the previous section for the Green functions. The
self -- energy potentials are now obtained by averaging
the Green functions Eqs.~\eqref{zeta} -- \eqref{K}
over the trajectory direction:
$\check\Sigma_T = i\Gamma\left<\check g_S\right>$.
This averaging does not affect,
of course, the induced gap function \eqref{fg-asimpt_gen}
 outside the primary vortex
core and, thus, the spectrum $\epsilon_{2}$ survives the
influence of the tunnel barrier disorder at least for $b>\xi_S$.
On the contrary, the subgap branches localized within the primary
vortex core are completely destroyed. Such
dramatic consequence of the momentum scattering is caused by the
averaging of electronic wave functions with different impact
parameters and consequent loss of
information about the CdGM states of the primary vortex.
A natural consequence of the momentum scattering is the
appearance of a finite broadening of energy levels for
trajectories with small impact parameters $b\lesssim\xi_S$.
Matching the solutions in the core and at large distances
gives the 
expression for the coefficient $C$ for $b\lesssim\xi_S$
and $|\epsilon|\ll\Gamma$,
\begin{multline}
 C\left[ \epsilon -\epsilon_2(b) + \frac{2\sqrt{\Gamma^2-\epsilon^2}}{\hbar v_{2F}} \int_0^{\infty} \Sigma_1\, ds
 -\frac{2\Gamma}{ \hbar v_{2F}}  \right. \\ \left. \times\int_0^{\infty} \Sigma_I^{loc} \,
ds\right] =\left[ \Gamma -\frac{2\epsilon}{ \hbar v_{2F}}
\int_0^{\infty} \Sigma_I^{loc} \, ds\right] \quad
\label{eqC-incoh}
\end{multline}
Since $|\Sigma_1| \sim
|\Sigma_I^{loc}| \sim \Gamma $ the pole of the coefficient $C$ is located at small energies $\epsilon \lesssim
\Gamma^2/\Delta\ll \Gamma$. Thus, for $\epsilon \ll \Gamma$ the expression for this coefficient takes the form
\begin{equation}
C\left[ \epsilon -\epsilon_2(b) + \frac{2}{ \xi_{2D}}
\int_0^{\infty}\left( \Sigma_1-\Sigma_I^{loc} \right) \, ds\right]
=\Gamma \ . \quad  \label{eqC-incoh-1}
\end{equation}
The localized self energies $\Sigma_1$ and  $\Sigma_I^{loc}$ can be
neglected for $\epsilon \sim \Gamma$. They also vanish for $|b|\gg
\xi_S$. In both these limits, Eq.~\eqref{eqC-incoh} transforms
into Eq.~\eqref{const2}. The integral term in the
 equation above can be written in terms of its real ${\beta(b)=\beta_I(b)-\beta_1(b)}$ and
 imaginary ${\gamma(b)=\gamma_I(b)-\gamma_1(b)}$ parts
as follows:
\begin{equation}\label{beta-gamma-def}
\frac{2}{\xi_{2D}}\int_0^{\infty}\left(
\Sigma_1-\Sigma_I^{loc} \right) \, ds=-\beta(b)\pm
i\gamma(b) \ .
\end{equation}
Here upper (lower) sign corresponds to the retarded (advanced) Green function.
Further we calculate the terms of real $\beta_{1,I}$ and imaginary $\gamma_{1,I}$ parts of the integral \eqref{beta-gamma-def},
 which are defined by the following expressions
\begin{equation*}
\beta_\alpha(b)=\frac{2}{\xi_{2D}}\int\limits_0^\infty\Re\Sigma_\alpha(s)ds
\ , \quad
\gamma_\alpha(b)=\frac{2}{\xi_{2D}}\int\limits_0^\infty\Im\Sigma_\alpha(s)ds
\end{equation*}
and play the role of energy shifting and spectral branch
broadening, respectively:
\begin{equation}
N_\epsilon(s,b)= \frac{\Gamma \gamma(b)e^{-|s|/\xi_{2D}}}{[\epsilon-\epsilon_2(b)- \beta(b)]^2
+\gamma^2(b)} \ ,\label{ARDOS}
\end{equation}
Since parameters $\beta, \gamma \sim \Gamma/\Delta$ and $\epsilon_2(b)/\Gamma \ll1$ are small
 for $b\ll \xi_{2D}$ and $|\epsilon|>\Gamma$, the LDOS reaches its bulk value in this region:
\begin{equation}
N(\rho,\epsilon)=\frac{\sqrt{\epsilon^2-\Gamma^2}}{|\epsilon|} \ .
\label{dos3}
\end{equation}
\begin{figure}[t]
\includegraphics[width=0.8\linewidth]{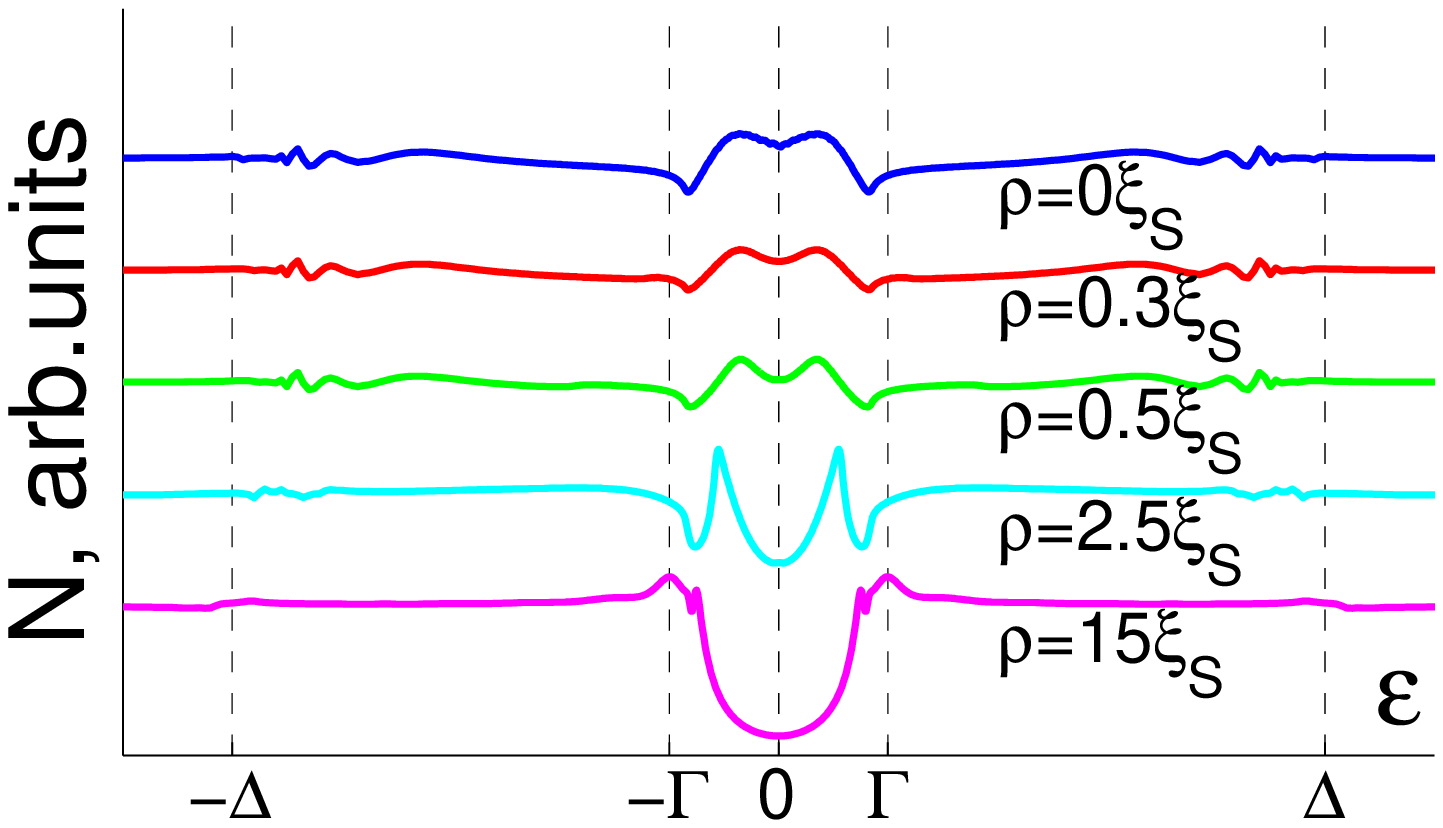}
\caption{(Color online) LDOS in logarithmic scale for incoherent tunneling in the
clean limit. Curves, taken for different distances $\rho$ from the vortex center,
are vertically shifted for clarity. The peaks in LDOS exist up to distances $\sim \xi_{2D}$.
Here $\Delta/\Gamma=5$, $q_v=1$.}
\label{fig-DOS-Incoh}
\end{figure}

Skipping the standard calculations of the above-defined integrals \eqref{beta-gamma-def}, we get the final expressions for parameters (see Appendix~\ref{app-beta-gamma} for details):
\begin{eqnarray}\label{gamma-beta-res}
\beta  = \left< \!\frac{\Gamma^2\pi q_v
}{Q\Omega}{\rm sign}(\epsilon+\Omega b)\! \right>_z \! ,
\gamma   =  \left<\!
\frac{\Gamma^2q_v}{Q\Omega}\ln \frac{\Delta_\infty}{|\Omega b
+\epsilon|}\!\right>_z\! .
\end{eqnarray}
The angular brackets denote averaging over the momentum $Q_z$
along the vortex axis in bulk, $\Omega =
\partial\epsilon_0/\partial b$. The DOS has a peak of height $\Gamma/\gamma$ at an energy
${\epsilon=\epsilon_2(b)+\beta(b)}$ shifted from the standard bound
state level.
This shift results in splitting of the ZBA \cite{ZBA} (Fig.~\ref{fig-DOS-Incoh}). For calculations we use the numerical
procedure similar to that used earlier for the
coherent limit; the induced potentials were
averaged over the cylindrical Fermi surface  in the bulk.

\subsection{Multiple core.\ Dirty SC with clean 2D
layer.}

Smearing of the energy dependence of the induced
potentials caused by disorder becomes even stronger
if the bulk SC has a short mean free path: $\ell \ll \xi_S$. In
dirty limit, the momentum averaged retarded (advanced) Green
functions are parameterized as follows:
\begin{equation}\label{selfen-dirty}
\check g_S^{R(A)}(\rho)=\check \tau_3\sin\Theta^{R(A)}+\check
\tau_2\cos\Theta^{R(A)} e^{-i\check \tau_3\phi} \ .
\end{equation}
We put $\Theta^{R(A)}=\pm \Theta_1 +i \Theta_2$. The boundary conditions for $\rho \to 0$ are $g^{R(A)}\to \pm 1$, $f^{R(A)}, f^{\dagger R(A)} \to 0$ which requires $\Theta_1 \to \pi/2$, $\Theta_2 \to 0$.
At large distances
$
\Theta_1 \to 0$, $\tanh \Theta_2 \to -\epsilon /\Delta_\infty$ for $\epsilon < \Delta_\infty$ while  $
\Theta_1 \to \pi/2
$, $\tanh \Theta_2 \to - \Delta_\infty/\epsilon $ for $\epsilon > \Delta_\infty$.
Then, $\Theta_2=0$ for $\epsilon \ll \Delta_\infty$, and the Usadel equation becomes 
\cite{GorkovKopnin}
\begin{equation}
D_S\left[ \nabla^2 \Theta_1 + \frac{\sin (2\Theta_1)}{2\rho^2}  \right] -2\Delta_0\sin \Theta_1 =0 \ . \label{eqUsadel1}
\end{equation}
The solution of Eq.~\eqref{eqUsadel1}
 has been found in Ref.~\onlinecite{GorkovKopnin}: $\Theta_1(\rho)$  monotonously decays from $\pi /2$  at the origin down to
zero at $\rho\gg\xi_S$.
The Green functions \eqref{selfen-dirty} determine the
induced vortex potentials $\check\Sigma_T=i\Gamma\check g_S$.
\begin{figure}[t]
\includegraphics[width=0.8\linewidth]{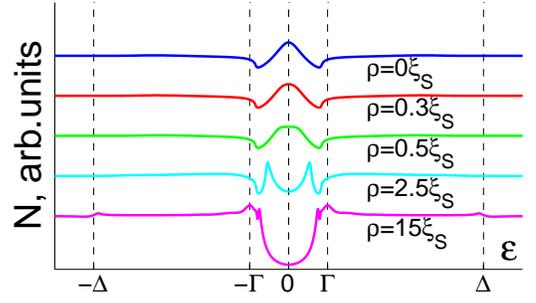}
\caption{(Color online) The local DOS in logarithmic scale for the dirty limit
with the parameters $\Delta/\Gamma=5$, $v_{2F}/V_F=1$. Curves, taken for different distances $\rho$ from the vortex center, are vertically shifted for clarity. }
\label{fig-DOS-dirty}
\end{figure}

For small impact parameter values
$b\ll \xi_S$ we get $\Sigma_I^{loc}=0$  and  the matching
condition takes the form:
\begin{equation}\label{bcond-theta2-dirty}
\xi_{2D} \theta(s_0) =2i\zeta(s_0) \int\limits_0^{\infty}
\sin\Theta\, ds
+2ig(s_0) b \ln [s_0/\xi_S] \ .
\end{equation}
The coefficient $C$ in this case has the only broadened pole at
$\epsilon=\epsilon_2(b)$:
\begin{equation}
C\left[ \epsilon -\epsilon_2(b) + i\gamma \right] =  \Gamma
\label{eqC-dirty} \ ,
\end{equation}
where the broadening
\begin{equation*}
\gamma = \frac{2\Gamma \sqrt{\Gamma^2-\epsilon^2}}{\hbar v_{2F}}
\int_0^{\infty} \sin\Theta \, ds \ ,
\end{equation*}
where the integral is taken along the trajectory.
For $|\epsilon| < \Gamma$ and $\rho<\xi_S$ 
the angle-resolved DOS can be written in the form
\begin{equation}
N_\epsilon(s,b)= \frac{\Gamma^2 }
{\sqrt{\Gamma^2-\epsilon^2}}\frac{\gamma(b)e^{-\lambda|s|}}{[\epsilon-\epsilon_2(b)]^2
+\gamma^2(b)} \label{DOS-dirty<} \ .
\end{equation}
Consequently, the LDOS has a peak of the height
$\sim\Gamma/\gamma(\rho)$ at energy $\epsilon=\epsilon_2(\rho)$.

For the energies above the induced gap $\epsilon >\Gamma$ and small impact parameter values $\epsilon _2(b),\gamma(b)\ll \Gamma$
the local DOS can be replaced by its bulk value:
\begin{equation}
N (\rho,\epsilon) = \frac{\sqrt{\epsilon^2 -\Gamma^2}}{|\epsilon|}
\label{DOS-dirty>}
\end{equation}

For $b\gg \xi_S$ the imaginary part  of energy decays
exponentially, and Eq.~\eqref{eqC-dirty} transforms into Eq.~\eqref{const2}.

The numerical
results shown in Fig.~\ref{fig-DOS-dirty} clearly demonstrate the broad peak in
the LDOS; this peak shifts and becomes sharper as the
distance from the vortex center increases.
For $\rho \gg \xi_S$, the LDOS approaches that obtained for the clean limit
in Figs~\ref{fig-DOS-Coher} and \ref{fig-DOS-Incoh}.
For calculations we
used the standard relaxation method \cite{RelaxMethod} of solving the Usadel
equation in the bulk and the Riccati parametrization for
Eilenberger equations in the 2D layer.

\subsection{Vortex core expansion.\ Dirty SC and
2D layer.}

To complete our analysis we briefly discuss the case of strong
disorder both in the bulk SC and in the 2D layer.
In this limit our model reduces to the one studied numerically in Ref.~\onlinecite{Golubov-vortex}.
The condition $\xi_S\ll \xi_{2D}=\sqrt{\hbar
D_{2D}/\Gamma}$ ensures that the short-distance
inhomogeneity in the induced vortex potentials inside the primary core region does not disturb the
adiabatic solution based on
Eq.~\eqref{fg-asimpt_gen}.  Indeed,
for momentum-orientation-averaged Green functions in 2D layer
$$\check g(\rho)=\begin{pmatrix}g_2 & f_2e^{i\phi}\\ -f_2^\dagger e^{-i\phi} & \bar{g}_2\end{pmatrix}=\int\frac{d^2k}{(2\pi)^2}\check g({\bf k, r})\ $$
one can derive the equation:
\begin{multline}
i D_{2D}\left[g_2(\nabla^2-\rho^{-2})f_2-f_2\nabla^2g_2\right]-\\
-2(\epsilon+\Sigma_1)f_2+2\tilde\Sigma_2g_2=0 \ ,
\end{multline}
with $\tilde\Sigma_2=\Sigma_2e^{-i\phi}$. This equation is similar to that derived by Kupriyanov \cite{Kupriyanov89} for a contact of two dirty superconductors.

Using a standard parametrization $\check
g(\rho)=\tau_3\sin\Psi+\tau_2\cos\Psi e^{-i\tau_3\phi}$ and the
expressions for the vortex potentials one can obtain
the following equation
\begin{equation}\label{eqUsadel-2D}
iD_{2D}\left[\nabla_\rho^2\Psi-\frac{\sin2\Psi}{2\rho^{2}}\right]-2\Gamma\sin(\Psi-\Theta)-2i\epsilon\cos\Psi=0 \ ,
\end{equation}
where $\nabla^2=\rho^{-1}\partial_\rho(\rho\partial_\rho)$ and
$D_{2D}=\hbar v_{2F}^2 \tau/2$~--~2D diffusion coefficient.
Integrating Eq.~\eqref{eqUsadel-2D}, multiplied by $\rho$, in a
small region around the origin (from $\rho=0$ to the value
$\xi_S\ll\rho_0\ll\xi_{2D}$) we find the matching condition for
the adiabatic Green function~(\ref{g2}, \ref{g-corr}):
\begin{multline}
D_{2D}\left[\left.\rho\frac{\partial}{\partial\rho}\Psi\right|_0^{\rho_0}+\int_0^{\rho_0}\frac{\sin2\Psi}{2\rho}d\rho\right]-\\
-2\int_0^{\rho_0}\rho d\rho\left[\Gamma\sin(\Psi-\Theta)+i\epsilon\cos\Psi\right]=0 \ .
\end{multline}

Considering the expansion $\Psi(\rho_0)=\Psi_0-K\rho_0$ with
$K={\partial\Psi(\rho_0)}/{\partial\rho}\sim\xi_{2D}^{-1}$ and
assuming $\Psi_0\neq\pi/2$ one obtains
${\cos\Psi_0\approx{\rho_0^2}/{(\xi_{2D}^2\ln\left({\rho_0}/{\xi_S}\right))}\ll 1} $.
This estimate confirms the conclusion that
the LDOS in the dirty limit follows the bulk LDOS
pattern scaled with the 2D coherence length $\xi_{2D}$ to within the second order terms in the small parameter
$\rho_0/\xi_{2D}$.

The resulting problem at low energies $\epsilon\ll\Delta_\infty$  coincides with
 that describing a standard vortex in a dirty SC \cite{Golubov} with the
gap value $\Gamma$. Thus, the full disordered system should reveal
the same LDOS patterns as in the bulk case, though scaled with the
much larger coherence length $\xi_{2D}$ instead of $\xi_S$.
This vortex-core expansion can account for
anomalously large vortex images observed in $MgB_2$
\cite{eskildsen} and in high-$T_c$ cuprates \cite{yeh}.

\section{Discussion.}\label{sec-discuss}
%

The results described above imply that the electronic states in the induced superconducting configurations strongly depend on the tunneling mechanism and on the crystal structure of bulk and 2D materials. The structure and symmetry of electronic states can be essentially different from those in the bulk SC. This imposes severe restrictions on possible realization of various exotic proximity electronic states\cite{Oreg10,Perfetto13} including Majorana states\cite{TI_S_prox_Majorana} and, in particular, Majorana states in the vortex cores. Our results directly show that the existence of zero-energy states in the proximity induced vortex core crucially depends on the tunneling mechanism underlying the proximity coupling between the 2D layer and bulk SC. One expects that the zero energy core state can exist for coherent tunneling between SC and 2D layer both having isotropic Fermi surfaces, provided the symmetry of the induced superconducting order permits.


It is known that a zero energy core state exists for a vortex with an odd vorticity in graphene monolayer with intrinsic superconductivity\cite{JackiwRossi81,bergman,Khaymovich-etal-09}. The graphene monolayer with proximity-induced superconductivity thus would seems to be a good candidate to look for a zero energy state.
However, the Fermi surface of graphene is highly anisotropic; it lies near the Dirac corners of the Brillouin zone with the group velocity directed radially from the Dirac points. This group velocity direction does not coincide with the direction of the Fermi momentum and of the Fermi velocity in the bulk SC as shown in Fig.~\ref{fig-anisotr}. Though the results of the previous sections were obtained within the quasiclassical approximation, they still can shed a light on the possibility of the zero energy state in graphene, especially for sufficient doping level when the quasiclassical approximation for graphene is justified \cite{Khaymovich-etal-09}. In this case the results of Sec.\ref{sec-anisotropic} can be applied. They show that each state in the induced vortex core with energy $\epsilon$ is coupled to an infinite set of levels in the primary core. It is the integral $I_{odd}$ which accounts for these states. Its real part deals with off-resonance states with eigen-energies not equal to $\epsilon$, while the imaginary part comes from  the resonance state with the same eigen-energy $\epsilon$. According to Sec.\ref{sec-anisotropic}, the real part of the integral $I_{odd}$ disappears for $\epsilon=0$ and $b=0$. The fate of the imaginary part depends on that is the zero energy in resonance with any state in the primary core or not. It is known that for an s-wave clean bulk superconductor the core levels are discrete with a minigap $\omega_0 \sim \Delta^2/E_F$ and no one lies at zero energy. Therefore, if the levels in the bulk are not broadened by disorder of by inelastic scattering, the imaginary part of $I_{odd}$ does not appear, and the zero-energy state seems to be intact. The discrete nature of the core states is, of course, beyond the quasiclassical approximation. Therefore, the above consideration gives only a hint towards the possibility of zero energy state. The detailed analysis is needed which would be based on the strict quantum mechanical description.
Note that an alternative possibility to save the zero energy states introducing a cylindrical cavity in the bulk superconductor
 has been considered in Refs.\cite{Iosel_Feig,Rakhmanov11}.


Other important feature of induced superconductivity in a LD system is
an extremely large coherence length $\xi_{2D}$. It provides a unique possibility to realize vortex configurations with quite unusual parameters. Here we discuss briefly some configurations which are of interest. The detailed analysis of all these situations requires special considerations which we postpone to future work. First of all we note that the results of Sections \ref{sec-ind-potentials}, \ref{sec-ind-vort} and the following sections are valid for $\xi_{2D}\ll {\rm min}(r_v,\lambda_L)$ where $r_v$ is the intervortex distance and $\lambda_L$ is the London penetration length in bulk SC.  If the vortex lattice in the bulk SC is dense enough with the intervortex distance $\xi_{2D} \lesssim r_v\ll \lambda_L$, the induced 2D vortex cores may start to overlap. The spectrum $\epsilon_2$ will then be modified due to intervortex
tunneling of QPs (see Ref. \cite{Meln_Sil_intervortex}).
The effect of the intervortex QP tunneling should be important provided
the splitting of the quantized energy levels due to this tunneling exceeds the minigap value.
The splitting can be estimated as $\Gamma \exp[-r_v/\xi_{2D}]$ while the minigap inside the induced vortex core
is of the order of $\Gamma^2/\hbar v_{2F}k_{2F}$. Thus, the ratio determining the intervortex tunneling efficiency is the exponent with a big prefactor, $\hbar v_{2F}k_{2F}\Gamma^{-1}\exp[-r_v/\xi_{2D}]$. It is this ratio which controls the interplay between the velocity of the trajectory
precession and QP tunneling speed. The changes in the QP spectrum become essential when $r_v\lesssim\xi_{2D}\ln(\hbar v_{2F}k_{2F}/\Gamma)$. The minigap in this case should vanish according to the analysis in Ref.~\onlinecite{Meln_Sil_intervortex}.

In some cases the 2D coherence length $\xi_{2D}$ can exceed the London penetration depth
$\lambda_L$; this depends on the properties of bulk SC and on the tunneling rate $\Gamma$. If $\xi_{2D}, r_v\gg \lambda_L$ the superconducting velocity vanishes along the
trajectories with $b>\lambda_L$, thus the spectral branch $\epsilon_2(b)$ saturates already for $b\sim\lambda_L$.

Our results for coherent tunneling can be directly generalized for clean $d$-wave bulk SCs with isotropic Fermi surfaces.
However, the incoherent tunneling destroys the superconducting coherence in the 2D layer. As a result,
the branch $\epsilon_2$ disappears, while the QP states for $\epsilon <\Delta$ have finite lifetimes for distances close to the vortex cores in bulk SC.

Considering possible experimental realizations of the induced vortex states one has to remember of the finite dimensions $L$ of the 2D layer. The large size of the induced vortex cores can lead to the situation typical for mesoscopic superconducting samples when $L$ is close to several $\xi_{2D}$'s. The criterion when the vortex spectrum transformation caused by the boundary effects in such systems becomes important
can be found using the results of Ref.~\onlinecite{Meln_Ryzh_Sil_mesa}. One only needs to replace the gap, the coherence length and the minigap by the appropriate values in the 2D layer. The criterion
appears to be very similar to that describing the efficiency of intervortex tunneling: the mesoscopic fluctuations of quantum levels in the 2D core become comparable with the minigap for $L\lesssim\xi_{2D}\ln(\hbar v_{2F}k_{2F}/\Gamma)$.

In conclusion, the model of proximity coupled 2D layer gives the possibility to study theoretically many spatially inhomogeneous situations including various configurations of induced vortices. Based on this model we have presented here description of the vortex core states for some typical tunneling mechanisms. In particular, our results can be used for interpreting the STM data on the vortex LDOS in superconductors through the model of a thin proximity layer present
at the surface of the bulk SC. Effect of a thin non-superconducting proximity layer can explain various experimentally observed features of the vortex LDOS and reveals that STM technique alone is not sufficient for identifying multicomponent or anisotropic energy gap.

\acknowledgements

We thank A.~Buzdin, G.~Volovik and A.~Smirnov for stimulating discussions.
This work was supported in part by EU 7th Framework Programme (FP7/2007-2013, Grant No. 228464 Microkelvin)
and by the Academy of Finland though its LTQ CoE grant
(project no. 250280), by the Russian Foundation for Basic Research,
by the Program ``Quantum Physics of Condensed Matter'' of the Russian Academy of
Sciences, the Russian president foundation (SP-
1491.2012.5),
 and by FTP ``Scientific and educational personnel of innovative Russia in
2009-2013''.

\appendix

\section{Calculation of self energies for incoherent tunneling}\label{app-beta-gamma}
 Assuming
small impact parameter values $b\ll\xi_S$, i.e., we calculate in this Appendix
the following integrals from the main text:
\begin{equation*}
\beta_\alpha(b)=\frac{2}{\xi_{2D}}\int\limits_0^\infty\Re\Sigma_\alpha(s)ds
\ , \quad
\gamma_\alpha(b)=\frac{2}{\xi_{2D}}\int\limits_0^\infty\Im\Sigma_\alpha(s)ds \ .
\end{equation*}
For this purpose we consider the case of the small impact parameter values $b\ll \xi_S$:
\begin{multline*}
\beta_I(b) =\frac{2\Gamma^2 b}{v_{2F}}\int_0^\infty \left< \frac{v_\parallel e^{-K}}{2Q\Omega\rho^2 }\right. \\
\times \left. \left[
1-\Re\frac{|\epsilon|}{\sqrt{\epsilon^2-\Omega^2\rho^2}}\right] \right>_z\, ds \ ,
\end{multline*}
where $\rho^2 =b^2+s^2$. In this case the first term in the above
integral is determined by $s \sim b$:
\begin{multline*}
\Gamma b\int_0^\infty \left< \frac{v_\parallel e^{-K}}{Q\Omega\rho^2 }
\right>_z\, ds=
\Gamma b\int_0^\infty \left< \frac{v_\parallel }{Q\Omega(s^2+b^2) } \right>_z\, ds\\
={\rm sign}(b)\Gamma \left< \frac{\pi v_\parallel }{2Q\Omega}
\right>_z \ .
\end{multline*}
The second one is determined by very small impact parameters and
reads:
\begin{equation*}
\int_0^{b_0}\frac{ds}{\sqrt{b_0^2-s^2}}=\frac{\pi}{2}\ , \;
\int_0^{b_0}\frac{ds}{(s^2+b_0^2)\sqrt{b_0^2-s^2}}=\frac{\pi
\Omega}{2|b\epsilon|} \ ,
\end{equation*}
where $b_0^2=\epsilon^2/\Omega^2 -b^2>0$. As a result, we find:
\begin{equation*}
\beta_I(b) ={\rm sign}(b) \frac{\Gamma^2}{v_{2F}} \left< \frac{\pi
v_\parallel }{Q\Omega} \chi(\Omega^2b^2-\epsilon^2) \right>_z \ ,
\end{equation*}
\begin{equation*}
\beta_1(b) =-{\rm sign}(\epsilon)\frac{\Gamma^2}{v_{2F}}\left<
\frac{\pi v_\parallel }{Q\Omega}\chi(\epsilon^2-\Omega^2b^2)
\right>_z \ . \label{real1}
\end{equation*}
Here $\chi(x)$ is the Heaviside theta-function, i.e., $\chi(x)=1$ for $x>0$ and  $\chi(x)=0$ for $x<0$.

After simplifying the expression for
$\beta(b)=\beta_I(b)-\beta_1(b)$ we obtain \eqref{gamma-beta-res}. For
$b\gtrsim \xi_S$ the quantity $\beta(b)$ decays as $\exp
(-2b/\xi_S)$.

The expressions for imaginary parts hold for any distances $\rho$
because the delta functions in the integrals
select only the trajectories that pass at small impact parameters:
\begin{multline*}
\gamma_1(b) =\frac{\Gamma^2}{v_{2F}} \int_0^\infty
\left<\frac{v_\parallel e^{-K}}{Q\sqrt{\Omega^2\rho^2 -\epsilon^2}}\chi (\Omega^2\rho^2 -\epsilon^2)\right>_z\, ds\\
=\frac{\Gamma^2}{v_{2F}} \left<\frac{v_\parallel
}{Q\Omega}\ln\frac{\Delta_\infty}{\sqrt{|\Omega^2b^2-\epsilon^2|}}\right>_z
\ ,
\end{multline*}
\begin{multline*}
\gamma_I(b) =\frac{\Gamma^2 b}{v_{2F}} \int_0^\infty
\left<\frac{\epsilon}{\Omega \rho^2}\frac{v_\parallel e^{-K}}
{Q\sqrt{\Omega^2\rho^2 -\epsilon^2}} \chi (\Omega^2\rho^2 -\epsilon^2)\right>_z\, ds\\
={\rm sign}(b\epsilon)\frac{\Gamma^2}{v_{2F}} \left<\frac{v_\parallel
}{Q\Omega}\ln\frac{\Omega|b|+|\epsilon|}
{\sqrt{|\Omega^2b^2-\epsilon^2|}}\right>_z \ .
\end{multline*}
Here we use the following expressions for the standard integrals:
\begin{equation*}
\int_{b_0}^{s_{max}}\frac{ds}{\sqrt{s^2\pm b_0^2}}=
\ln\frac{\Delta}{\sqrt{|\Omega^2b^2-\epsilon^2|}} \ ,
\end{equation*}
where $s_{max}\sim \xi_S$, and
\begin{equation*}
\int_{b_0}^{s_{max}}\frac{ds}{\sqrt{s^2\pm b_0^2}(s^2+b^2)}=
\frac{\Omega}{|b\epsilon|}\ln\frac{\Omega |b|+ |\epsilon|
}{\sqrt{|\Omega^2b^2-\epsilon^2|}} \ .
\end{equation*}
The imaginary terms also decay exponentially for $b\gtrsim \xi_S$.
The expression for $\gamma(b)=\gamma_1(b)-\gamma_I(b)$ gives
\eqref{gamma-beta-res}.

\end{document}